\documentclass[useAMS,usenatbib]{mn2e}
\usepackage{graphicx}
\usepackage{epsfig}
\usepackage{multirow}
\usepackage{float}

\title[The Best Constraints on A Super-Eddington Accretion Flow] {The Best
  Constraints on A Super-Eddington Accretion Flow: XMM-Newton Observations of An 
Intermediate-mass Black Hole}

\author[C. Jin, C. Done, M. Ward]
{Chichuan Jin$^{1}$, Chris Done$^2$, Martin Ward$^{2}$\\
$^{1}$Qian Xuesen Laboratory of Space Technology, China Academy of Space Technology, Youyi Road, Beijing, 
100094, China\\
$^{2}$Department of Physics, University of Durham, South Road, Durham, DH1 3LE, UK
}

\begin{document}

\date{}

\maketitle

\label{firstpage}

\begin{abstract}
RX J1140.1+0307 (hereafter RX1140) is a Narrow Line Seyfert 1 (NLS1)
with one of the lowest black hole masses known in an AGN (M~$\la
10^6$M$_{\sun}$).  We show results from two new {\it XMM-Newton}
observations, showing soft 2-10~keV spectra, a strong excess at lower
energies, and fast X-ray variability as is typical of this class. The
soft excess can be equally well fit by either low temperature
Comptonisation or highly smeared, ionised reflection models, but we
use a covariance analysis of the fast X-ray variability as well as lag
and coherence spectra to show that the low temperature Comptonisation
model gives a better description of the break in variability
properties between soft and hard X-rays.  Both models also require an
additional component at the softest energies, as expected from the
accretion disc. However, this inner disc spectrum does not join
smoothly onto the variable optical and far UV emission (which should
be produced in the outer disc) unless the mass is underestimated by an
order of magnitude. The variable optical and far UV emission instead
suggests that $L/L_{Edd}\sim 10$ through the outer disc, in which case
advection and/or wind losses are required to explain the observed
broadband spectral energy distribution. However, the similarity of the
X-ray properties of RX1140 to other simple NLS1 such as PG
1244+026, RE J1034+396 and RX~J0136 means it is likely that these are
also super-Eddington sources. This means their spectral energy
distribution cannot be used to determine black hole spin despite
appearing to be disc dominated. It also means that the accretion
geometry close to the black hole is unlikely to be a flat disc as
assumed in the new X-ray reverberation mapping techniques.

\end{abstract}


\section{Introduction}
\label{sec-introduction}

Intermediate mass black holes (IMBH), which are taken to refer to
masses from $10^{4}-10^{6}M_{\odot}$, are of particular interest
because this population would fall within the gap between the two
principal black hole populations, i.e. super-massive black holes (SMBH)
with $M\ge10^{6}M_{\odot}$ and stellar mass black holes with
$M\sim10M_{\odot}$. Currently there are very few well studied IMBH,
two examples are NGC 4395 (Filippenko \& Ho 2003) and POX 52 (Kunth,
Sargent \& Bothun 1987; Barth et al. 2004). 

For over two decades, the most robust technique to measure black hole
masses in Active Galactic Nuclei (AGN) has been the reverberation
mapping (RM) technique, which estimates the radius of the broad line
region (BLR) by measuring the time-lag between the line emission from
the BLR and the intrinsic continuum emission from the accretion
disc. Since optical RM requires well sampled spectroscopical
monitoring of variable AGN for months or years, the total number of
sources with RM masses is currently less than forty (e.g. Kaspi et
al. 2000; Peterson et al. 2004, 2005; Denney et al. 2006, Bentz et
al. 2009b; Du et al. 2014). Instead, the RM measurements can be used
to derive a radius-luminosity relation which can then give black hole
mass from single epoch measurements using the Balmer line width
(normally that of H$\beta$ FWHM) and luminosity (generally
monochromatic luminosity at 5100\AA, hereafter ${\lambda}L_{5100}$
e.g. Bentz et al. 2006, 2009a; Denney et al. 2010).  Employing this
method, Greene \& Ho (2004) identified 19 IMBH in the SDSS
database. Later on Dong et al. (2007) and Greene \& Ho (2007) extended
the IMBH sample to include several hundred sources.

RX~J1140.1+0307 (hereafter RX1140), also referred to as GH~08 or SDSS
J114008.71+030711.4, is amongst the original 19 IMBH sample of Greene
\& Ho (2004). It has a redshift of 0.081 (co-moving distance of
336~Mpc). The SDSS and {\it HST} images show a resolved disc
component, plus a bar and AGN components (Greene, Ho \& Barth 2008). The
H$\beta$ FWHM measured from the SDSS spectrum is $700-780~km~s^{-1}$
(Ai et al. 2011; Jin et al. 2012a). However, to get mass also requires
${\lambda}L_{5100}$, and this is made more difficult by the strong
host galaxy contamination. Deconvolution of the HST image gives an
estimated AGN continuum of ${\lambda}L_{5100}\sim 6\times
10^{42}erg~s^{-1}$, which is somewhat smaller than the observed
variable flux of ${\lambda}L_{5100}\sim 8\times 10^{42}erg~s^{-1}$
(Rafter et al. 2011). This gives a black hole mass of $8-10\times
10^{5}M_{\odot}$ (Greene, Ho \& Barth 2008; Ai et al. 2011). However,
Rafter et al. (2011) applied the RM technique to RX1140, and reported
an upper limit of 6 light days for $R_{BLR}$, which gives
$M~{\la}~5.8{\times}10^{5}M_{\odot}$, but this is probably too strict
a limit given that their average sampling time of 6.1 days. A more
conservative upper limit from the reverberation is probably a factor 2
larger, giving $M\la 1.2\times 10^6M_{\odot}$, consistent with the
$H\beta$ mass estimate.

For this small a black hole mass, the total AGN luminosity is likely to
be a high fraction of Eddington, with Zhang \& Wang (2006) estimating
$L/L_{Edd}\sim 1.6$ from $L_{bol}=9\times {\lambda}L_{5100}$ (though
using the luminosity corrected for host galaxy contamination reduces
it to $\sim 0.6$).  So far, such high Eddington fraction accretion
flows, with $L/L_{Edd}\ga 1$, are found in the subset of broad line
AGN known as Narrow-Line Seyfert 1 (e.g. 
Osterbrock \& Pogge 1985; Boroson \& Green 1992;  Leighly 1999). Their narrow Balmer
lines indicate relatively small black hole masses ($10^{6-7}M_\odot$:
Boroson 2002), so that their bolometric
luminosities are close to Eddington. For such high Eddington fractions
it is clearly a moot point as to whether the H$\beta$ scaling
relations hold. Marconi et al. (2009) suggest a correction to the mass
scalings such that the BLR clouds trace the effective gravity, but the
lack of observed wind features in the low ionisation BLR lines form a
strong argument against such effects being important. This could
indicate that the clouds are optically thick to electron scattering,
with columns of $>10^{24}$~cm$^{-2}$, or that the radiative
acceleration only affects the front face of the cloud (Baskin, Laor \&
Stern 2014). Thus it seems probable that the H$\beta$ line widths
reflect the black hole mass even at Eddington and potentially beyond.

One of the ubiquitous features of high Eddington fraction AGN is their
prominent soft X-ray excess superposed on a steep X-ray power law
which is very variable. These properties are seen in the X-ray
spectrum of RX1140 (Miniutti et al. 2009; Ai et al.
2011). While the variability scales with black hole mass, making it
unsurprising that the X-ray variability of RX1140 is amongst
the strongest seen AGN (e.g. Ponti et al. 2012; Ludlam et al. 2015), the
origin of the soft excess is more puzzling. Models which reproduce the
shape include optically thick, low temperature Comptonisation
(e.g. Laor et al. 1997; Magdziarz et al. 1998; Gierli\'{n}ski \& Done
2004; Jin et al. 2013), ionised, highly smeared reflection
(e.g. Fabian \& Miniutti 2005; Zoghbi et al. 2010; Fabian et al. 2013;
Fabian, Kara \& Parker 2014; Uttley et al. 2014), and smeared
absorption (Gierli\'{n}ski \& Done 2004; 2006).  The smeared
absorption model is disfavoured on theoretical (Schurch \& Done 2007)
as well as observational grounds (Miniutti et al. 2009, Ai et al. 2011),
but both Comptonisation and smeared reflection provide a good fit to
the soft excess in this object (Ai et al. 2011).

However, for such low mass, high mass accretion rate AGN, the disc
itself should also contribute a significant fraction of the emission
in the soft X-ray bandpass. It is well known that the accretion flow
spectrum is not well described by a disc in standard broad line
Seyfert 1s and Quasars (e.g. Elvis et al. 1994), but the higher
Eddington ratio NLS1 have spectra which do appear to be more
disc-dominated (Jin et al. 2012a,c; hereafter J12a; c; Done et al. 2012,
hereafter D12). This is most easily seen in the context of the full
Spectral Energy Distribution (SED) as the optical/UV disc
normalisation is determined by the product of mass and mass accretion
rate, with $L_\nu\propto (M\dot{M})^{2/3}$. Thus this directly
measures mass accretion rate if the black hole mass is known, and 
hence the total luminosity $L=\eta \dot{M}c^2$, where the efficiency
$\eta$ depends on black hole spin (e.g. Davis \& Laor 2011; Done et al.
2013). This also  predicts where the disc disc emission peaks, and
show that the disc spectrum should extend into the soft X-ray bandpass
for the NLS1 even at low spin (Done et al. 2013).

RX1140 is the lowest mass NLS1 included in the sample of 51
unobscured Type 1 AGN of J12a,c. These papers carried out a detailed
study of the broadband SED from
optical to hard X-rays, fitting them with the new accretion models
which allow some of the accretion energy to be dissipated in a
Comptonised soft X-ray excess and power law tail as well as in the
thin disc (D12). These assume energy conservation i.e. that there is
no strong energy loss via advection and/or winds as might be expected
in highly super-Eddington sources. J12a, c used the black hole mass
was a fit parameter within the uncertainties derived from using the
intermediate and broad H$\beta$ line component widths as lower and
upper limits, respectively. This gave a best fit value of
$2.9{\times}10^{6}M_{\odot}$ for this source (J12a), which increased
to $6{\times}10^{6}M_{\odot}$ when the updated SED models including a
colour temperature correction to the disc emission were used (D12,
J12c). Clearly the mass of RX1140 derived from broadband SED
fitting is much higher than the virial and RM mass estimates, and this
requires further consideration.

The X-ray properties of RX1140 are very similar to a small
group of the most extreme NLS1s, such as PG 1244+026, RE J1034+396 and
RX J0136.9-3510, which all exhibit strong X-ray variability, prominent
and featureless soft excesses and are all reported to be accreting
around the Eddington limit (Middleton et al. 2009; Jin et al. 2009;
Jin et al. 2013). In order to investigate the physical interpretation
for the soft excess seen in RX1140, a detailed timing
analysis must be performed in conjunction with the spectral analysis
(e.g. Jin et al. 2013). In achieve this we applied for a long {\it
XMM-Newton} observation for this source. This was successful, but was
split into two observations separated by two weeks. In this paper we
present the results from a combined analysis of these two new{\it
XMM-Newton} observations, plus the previous one extracted from the
archive.

This paper is organised as follows: Section~\ref{sec-data-reduction}
describes the data reduction procedures used for the {\it XMM-Newton}
observations. Section~\ref{sec-var} presents variability properties of RX1140.
Section~\ref{sec-specfit} presents a combined spectral and
variability study of the target. We compare two physical models for explaining the
properties of the 0.3-10 keV spectra and variability. In
Section~\ref{sec-sed}, we perform broadband SED fitting by including
ultraviolet (UV), optical and near-infrared data. In
section~\ref{sec-discussion} we discuss issues such as various
estimates of the black hole mass, UV/optical luminosity and the
mechanism responsible for the soft X-ray excess. Our summary and
conclusions are given in
section~\ref{sec-summary}. When converting flux to
luminosity, we adopt a flat universe model with the Hubble constant
H$_{0} = 72$ km s$^{-1}$ Mpc$^{-1}$, $\Omega_{M} = 0.27$ and
$\Omega_{\Lambda} = 0.73$. 

\section{Data Reduction}
\label{sec-data-reduction}

RX1140 was observed by the {\it XMM-Newton} satellite on 3rd,
Dec., 2005 (Principal Investigator: Dr. Giovanni Miniutti) (hereafter:
Obs-1). We proposed a longer observation with the aim to study its
X-ray variability. Our observation was divided into two parts due to
satellite scheduling. These were carried out on 18th, Dec., 2013
(hereafter: Obs-2) and 1st, Jan., 2014 (hereafter: Obs-3). During the
three observations, all EPIC cameras were in the full window mode. We
used {\tt SAS} v13.5.0 and the latest calibration files, and followed
the standard procedures to reduce the data. We chose the source
extraction region to be a circular region of radius 45\arcsec for each
EPIC camera. The background was selected from a nearby circular region
with the same radius as for the source.  For Obs-1 the net source
count rates are $0.62~ct~s^{-1}$, $0.13~ct~s^{-1}$ and
$0.13~ct~s^{-1}$ for the PN, MOS1 and MOS2, respectively.  Obs-2 has a
mean count rate that is half that of Obs-1. Obs-3 has a similar mean
count rate as Obs-1. All of these count rates are well below the
threshold count rates capable of causing a photon pile-up effect in
the full window mode of each EPIC camera. 

We selected data with {\tt{PATTERN} $\le$ 12} for MOS1 and MOS2, and
{\tt{PATTERN} $\le$ 4} for the PN. Light curves were extracted from
both the source and background regions. There were high background
flares in all three observations. Therefore we visually checked the
background light curves to identify high background sections and
removed them. Then the background was subtracted from the source light
curve using {\tt LCMATH} in {\tt FTOOLS}. Spectra were extracted for
the source and background regions, separately.  Response matrices were
produced using {\tt RMFGEN} and {\tt ARFGEN}.  Areas for the source
and background regions were calculated using {\tt BACKSCALE}.  We also
used the {\tt SAS} task {\tt RGSPROC} to extract the 1st order spectra
from RGS1 and RGS2. Response matrices for the RGS were produced with
{\tt RGSRMFGEN}.  All spectra were rebinned by {\tt GRPPHA} with a
minimum of 25 counts per bin, so that the $\chi^2$ fitting is
appropriate.But the signal-to-noise of RGS spectra are too low to provide
further information, so we do not present them in this paper.
All spectral fittings were performed in {\sc xspec} v12.8.2 (Arnaud 1996).

There are also simultaneous Optical/UV observations obtained using the
{\it XMM-Newton} Optical Monitor (OM). OBs-1 has UVW1 and UVM2 filter
data, while Obs-2 and 3 have U, B and UVW1.  We searched the OM source
list file to obtain the count rate of RX1140, in each
available filter, and inserted these values into the standard OM data
file template {\it om\_filter\_default.pi}. The data file was then
combined with the `canned' response
files\footnote{http://heasarc.gsfc.nasa.gov/FTP/xmm/data/responses/om}
to be ready for {\sc xspec} fitting.

\begin{figure}
\begin{center}
\includegraphics[bb=50 36 432 540, clip=,scale=0.62]{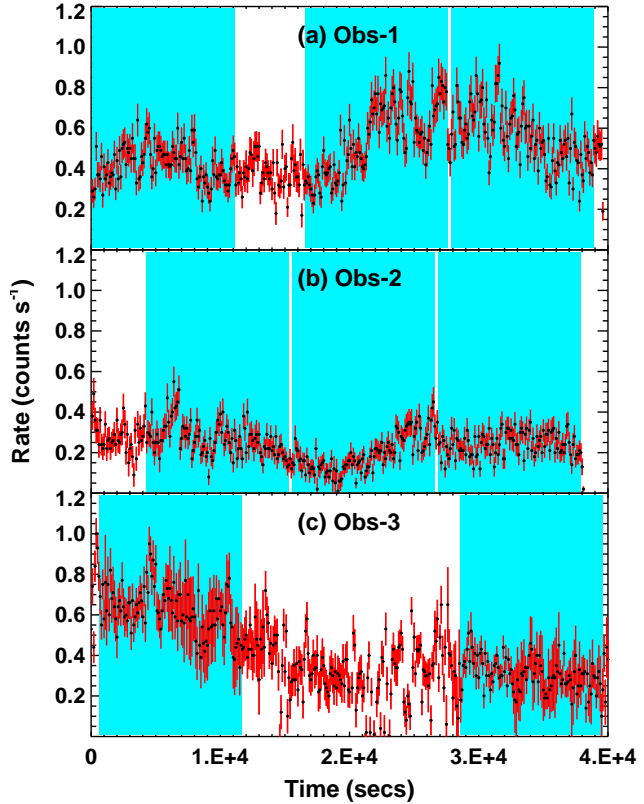}
\end{center}
\caption{The 100 s binned, background subtracted, 0.3-10 keV light curves for
  Obs-1, Obs-2 and Obs-3. Every cyan region is a 11 ks segment that
  is free from high background contaminations.}
\label{fig-lc}
\end{figure}

\section{Variability Study}
\label{sec-var}

\begin{figure*}
\centering
\includegraphics[angle=90,scale=0.6]{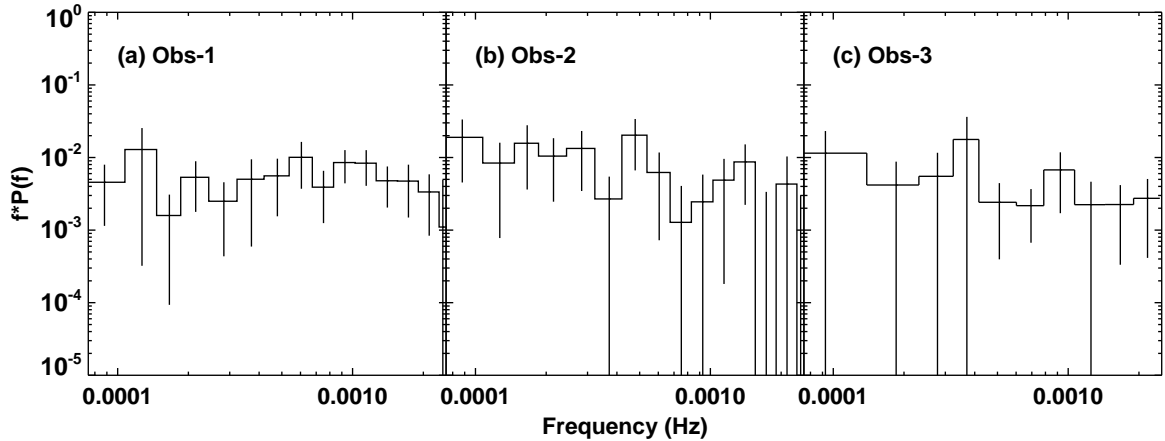}
\caption{Power Spectra Density for the 0.3-10~keV lightcurves of the 
three observations obtained using {\sc  FTOOL} {\tt powspec}
after subtracting the Poisson counting noise. For Obs-3, only the two 11 ks segments in Figure~\ref{fig-lc} were used to avoid the influence of high background.}
\label{fig-powspec}
\end{figure*}

The background subtracted PN light curves over the total energy band
(0.3-10~keV) are shown in Figure~\ref{fig-lc}. Strong variability is
observed over all timescales.  This is also confirmed by their Power
Spectral Density (PSD) in Figure~\ref{fig-powspec}. The flat PSD
indicate that intrinsic variability (i.e. Poisson noise subtracted)
exists throughout the 0.2-10 ks timescale. None of the
observations show any sign of a high frequency break. This is clearly
different to the case of PG 1244+026, where the 0.3-1~keV power spectrum 
drops substantially above $\sim 10^{-3}$~Hz (Jin et al. 2013), thus strongly suggests
that RX1140 is substantially lower mass than PG 1244+026, and that
the mass is $\la 10^6M_\odot$ (Ponti et al. 2012;
Kelly et al. 2013; Ludlam et al. 2015). 

The time-averaged X-ray spectra for Obs-1 (black) and Obs-2 (red) are
shown in the upper panel of Figure~\ref{fig-specvar}. The mean
spectrum of Obs-3 is almost identical to Obs-1, so it is not shown in
the figure. The Obs-1 and Obs-2 spectra are quite similar, but a ratio of the best fit
model to Obs-1 (black solid model), rescaled to the observed 2-10~keV
spectrum of Obs-2 (red solid model) under-predicts the soft emission.
This longer term (weeks to years) variability can be compared to the
rapid variability. The shapes of high ($ pn\_rate\ge0.7~ct~s^{-1}$: orange) and
low ($pn\_rate\ge0.35~ct~s^{-1}$: blue) count rate spectra from Obs-1
are identical within errors to the mean spectrum. This can be seen in
the lower panel of Figure~\ref{fig-specvar}, which gives the ratio of
each dataset to the mean spectral model renormalised to the 2-10~keV
count rate. 

\subsection{{\it RMS} Spectra}

Fractional variability observed in different energy bins, i.e. the
{\it RMS} spectra (e.g. Edelson et al. 2002; Markowitz, Edelson \&
Vaughan 2003; Vaughan et al. 2003), can be very useful in decomposing
the different spectral components. We use the segment algorithm
detailed in Wilkinson \& Uttley (2009) to calculate the {\it RMS}.
For every energy bin, the light curve is divided evenly into several
segments. Then the excess variance is calculated for every segment and
averaged over all segments. The square-root of the excess variance is
{\it RMS}. The error-bars are calculated using Equation B2 in Vaughan
et al. (2003). The binning time and segment length determine the
timescale for which the {\it RMS} is calculated. The high background
intervals interrupt the continuity of the light curve. In Obs-3, the
longest continuously sampled segment is only 11 ks, while it is longer
in Obs-1 and Obs-2. To mitigate bias from high background
contamination, we choose a longest segment length of 11 ks, and so are
able to identify a total of eight segments with little high background,
i.e. three segments in Obs-1, three in Obs-2 and two in Obs-3 (see the cyan
regions in Figure~\ref{fig-lc}).

\begin{figure*}
\includegraphics[scale=0.65,angle=90]{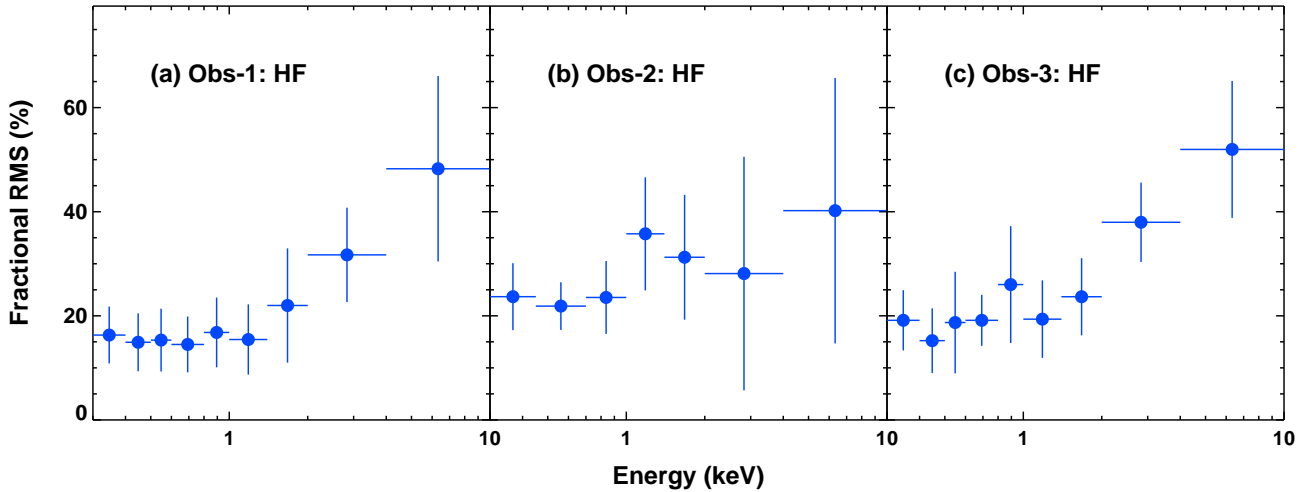}
\caption{The HF (0.2-2~ks) {\it RMS}
  spectra for the three observations. Obs-1 and Obs-3 have similar
fast spectral variability as well as similar mean spectra, while Obs-2
has more fast variability at low energies than both Obs-1 and Obs-3.}
\label{fig-rms}
\end{figure*}

There are insufficient data to produce frequency-resolved {\it RMS}
spectra. Instead, we calculate only {\it RMS}
of the highest frequency (HF) variability using timescales of 0.2-2 ks
from the eight 11ks segments.  The
results are shown in Figure~\ref{fig-rms}.  Interestingly, the HF {\it
RMS} spectra reveal a rising shape from soft to hard X-ray ranges in
Obs-1 and Obs-3, which is similar to that observed in some other high
mass accretion rate NLS1s, such as PG~1244+026 (Jin et al. 2013), RE
J1034+396 (Middleton et al. 2009) and RX~J0136.9-3510 (Jin et
al. 2009). This difference in variance between soft and hard energy
bands can be easily explained if the soft X-ray excess varies less
than the power law tail, favouring models where this is a true
additional component e.g. from soft Comptonisation
(e.g. Gierli\'{n}ski \& Done 2004, Middleton et al. 2009, Jin et
al. 2013, Matt et al. 2014) rather than due to ionised, smeared
reflection, as this tends to contribute equally in soft and hard bands
(Gierli\'{n}ski \& Done 2006).

\begin{figure}
\includegraphics[scale=0.48]{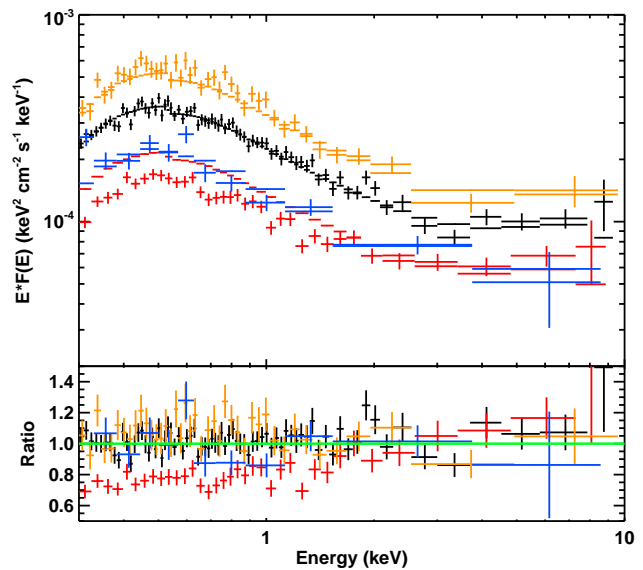}
\caption{Time-averaged spectra of Obs-1 (black) and Obs-2 (red)
together with the highest (orange) and lowest (blue) spectra seen
within Obs 1. The best-fit model of Obs-1 is the same as in
Figure~\ref{fig-specfit} (a) (black model), then renormalised to match the
2-10 keV of each spectrum. The lower panel shows the ratio of each
dataset to the model, showing significant spectral variability only in
Obs-2, where the soft X-rays are weaker.}
\label{fig-specvar}
\end{figure}

Another key feature seen in Figure~\ref{fig-rms} is that the 0.3-1 keV
HF {\it RMS} in Obs-2 is rather different to that of Obs-1 and Obs-3, with
much less difference between the fast variability as a function of
energy (more HF power below 1 keV, and marginally less at the highest
energies). Whatever happened to depress the soft spectrum relative to
the mean (Figure~\ref{fig-specvar}) may also increased its fast
variability (see Gardner \& Done 2015 for a possible explanation of
this in terms of cloud occultions in the inner disc).

\subsection{Covariance spectra}

The {\it RMS} variability spectrum shows all variability at each
energy, but where there may be multiple components. A more powerful
technique is the covariance
spectrum (developed by Wilkinson \& Uttley 2009) as it determines
the spectrum of the variability correlated with
a given energy band lightcurve used as a reference.
For the NLS1 PG 1244+026, Jin et al. (2013) showed that
using the fastest variability in the hard X-ray (2-10~keV) lightcurve
as a reference band gave strong evidence for a Comptonisation origin
for the soft X-ray excess, while using the fastest variability in the
0.3-1~keV band showed evidence for a contribution at the lowest
energies from the disc.

Here we apply the same technique to our datasets to produce the HF
(0.2-2~ks) covariance spectra using hard (2-10 keV) and soft
(0.3-1~keV) X-rays as the reference band. We use only PN data so as
not to have differences in response from combining PN and MOS. 
But the data quality means
that the signal-to-noise of the covariance spectra for RX1140 is much
lower than for PG 1244+026.
Then we study both the covariance spectra and the time-averaged spectrum 
(see Figure~\ref{fig-specfit}) in
order to try to determine the origin of the soft X-ray excess,
i.e. whether it can be better fit by an additional Comptonisation
model or by relativistically smeared reflection, and whether there is
also a contribution from the accretion disc itself. 

\section{Spectral Analysis}
\label{sec-specfit}

\begin{figure*}
\begin{tabular}{cc}
\includegraphics[scale=0.55]{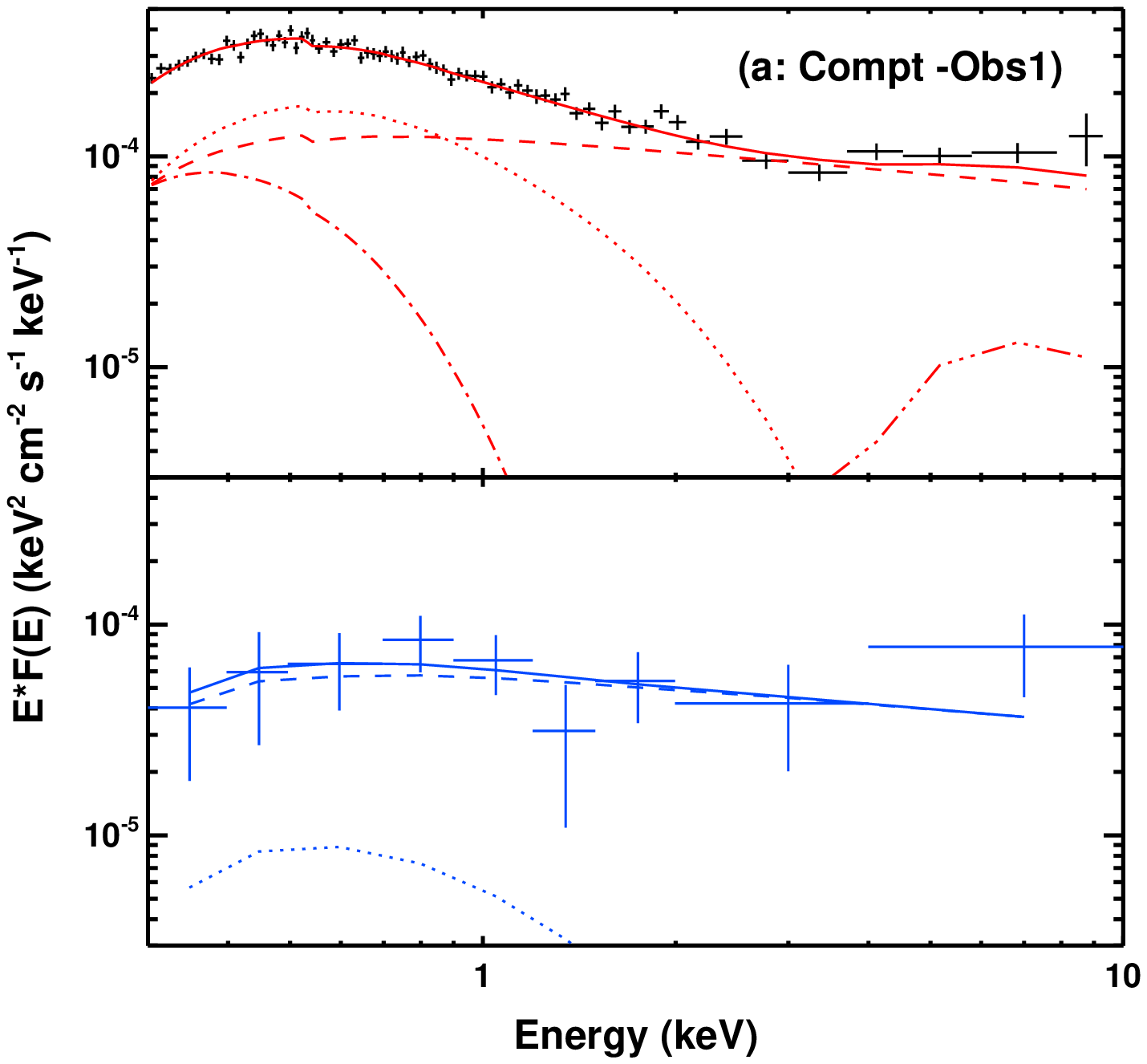} &
\includegraphics[scale=0.55]{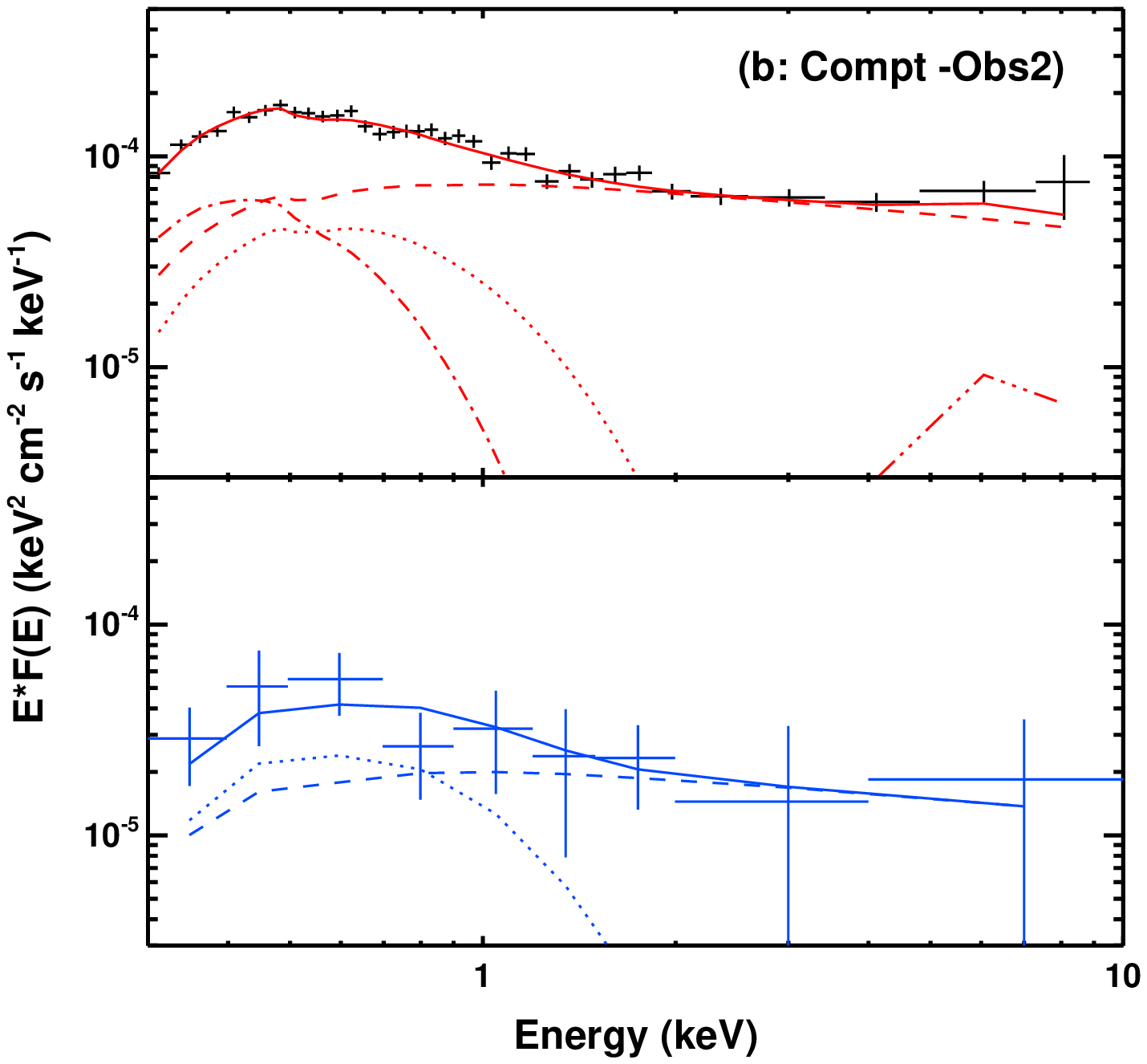} \\
\includegraphics[scale=0.55]{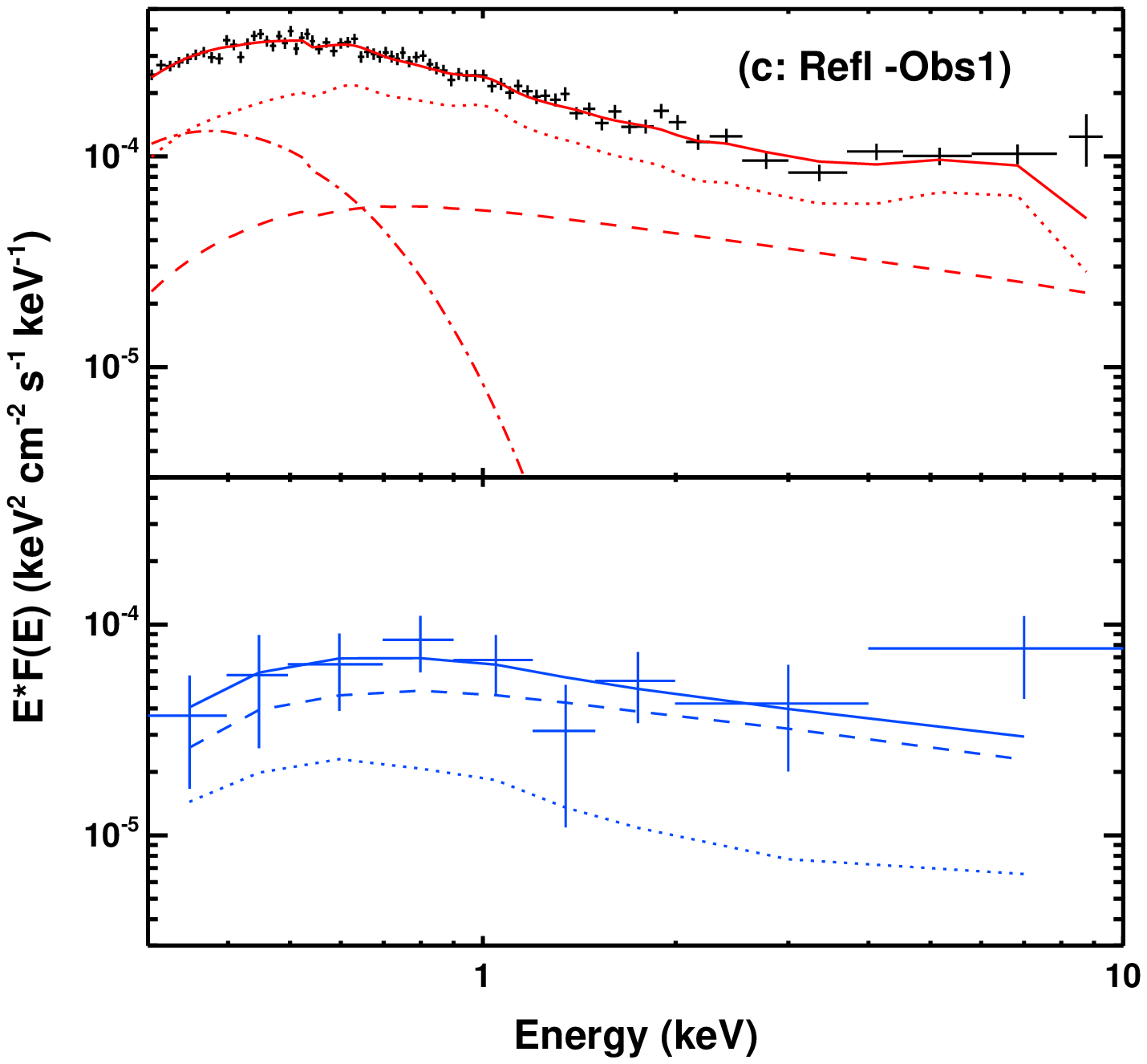} &
\includegraphics[scale=0.55]{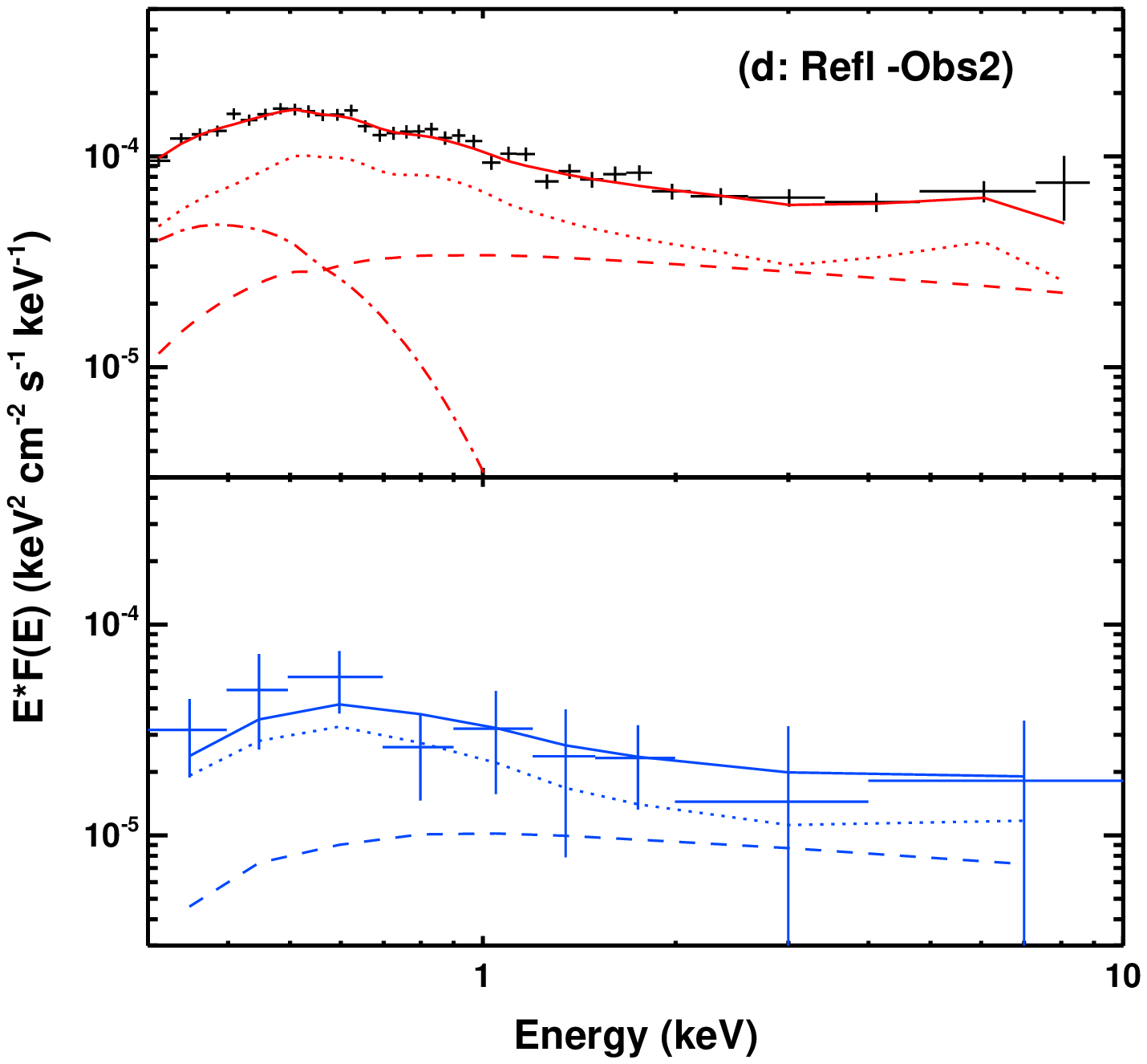} \\
\end{tabular}
\caption{X-ray spectral fitting of Obs-1 (Panel Compt-Obs1, Refl-Obs1)
  and Obs-2 (Panel Compt-Obs2, Refl-Obs2).  The upper row shows
  the results of Comptonisation fitting to the mean
  (red) and HF covariance (blue) spectra. In each panel, red solid line is the
  best-fit model, red dashed line is the hard X-ray Comptonisation component,
  red dotted line is the soft X-ray Comptonisation component, red dash-dot line
  is the hot accretion disc component. The blue lines are spectral fits to the HF
  covariance spectra with only the soft and hard Comptonisation components.
  The lower row panels present the results of reflection model fitting. The red
  dashed line is the intrinsic hard Comptonisation component, the red dotted
  line is the reflected component by accretion disc, the red dash-dot line is the
  accretion disc component. The blue lines are spectral fits to the covariance
  spectra with only the hard Comptonisation and reflection components.}
\label{fig-specfit}
\end{figure*}

The X-ray spectrum derived from Obs-1 has been analysed
previously. Miniutti et al. (2009) found that an X-ray reflection plus
a thermal disc model, gives a better fit than the smeared absorption
model. Ai et al. (2011) found that reflection and Comptonisation
models gave comparably good fits, and that both these were better 
than smeared absorption and $p$-free disk models. The shape
of the HF {\it RMS} and mean spectra in Figure~\ref{fig-specvar} and
Figure~\ref{fig-rms} also indicates that RX1140 may be similar to other
extreme NLS1s such as PG 1244+026, RE J1034+396 and RX J0136.9-3510,
whose soft X-ray excess can be well interpreted by Comptonisation
model. Therefore, we try both Comptonisation and reflection models for
Obs-1 and Obs-2 spectra.  The best-fit models are shown in Figure~\ref{fig-specfit}.

\subsection{Comptonisation Model}
\label{sec-comp}

For the Comptonisation scenario, we use a low temperature, optically
thick Comptonisation model ({\tt comptt} model, Titarchuk 1994; Hua \&
Titarchuk, 1995; Titarchuk \& Lyubarskij 1995) to fit the soft X-ray
excess; a high temperature Comptonisation ({\tt nthcomp}
model in {\sc xspec}, Zdziarski, Johnson \& Magdziarz 1996; {\.Z}ycki,
Done \& Smith 1999) to fit the power law tail, together with its
neutral, smeared reflection ({\tt kdblur*pexmon}, Laor 1991; recoded
as a convolution model, Nandra et al. 2007). We also include an
accretion disc component ({\tt diskbb} model, Mitsuda et al. 1984;
Makishima et al. 1986). Using the equation in Peterson (1997) and
assuming $r_{in}=6r_{g}$, $Log(M)=5.77$ and $L/L_{Edd}=2.69$ (Miniutti
et al. 2009), we calculate the temperature at the inner radius
(T$_{in}$) to be 110~eV. Including a colour temperature correction
would increase this, while allowing the some of the inner disc power
to be dissipated in powering the soft and hard X-ray Compton
components would decrease it, so we freeze the parameter T$_{in}$ at
this temperature. It also serves as the temperature of the seed
photons for both soft and hard X-ray Comptonisation components.  We
fix the Galactic column at $N_{H}~=~1.91{\times}10^{20}~cm^{-2}$
(Kalberla et al. 2005) but allow additional free absorption by the
host galaxy gas column ({\tt zwabs}, Morrison \& McCammon 1983).  The
complete {\sc xspec} model parameters are given in
Table~\ref{tab-specfit}.

The HF variability is likely to originate mainly from the inner region
of the flow, where the hard X-ray emission is produced. Hence we
expect that the spectrum correlated with the HF 2-10~keV variability
should have the same shape as the high temperature Comptonisation
component which dominates the 2-10~keV spectrum in this
model. However, reflection (and associated thermal reprocessing which
could contribute to the soft X-ray excess: Gardner \& Done 2015a)
should also vary in a correlated manner as the expected reverberation
timescale (Fabian et al. 2009) is short for a low mass black hole.
Only the disc is not expected to have any variability on these
timescales.

\begin{table*}
  \begin{minipage}{175mm}
   \centering
   \caption{The {\sc xspec} model and the best-fit
     parameters in Figure~\ref{fig-specfit}. The upper and lower limits are the
     90\% confidence range.}
     \begin{tabular}{lllll}
\hline
Model Name &   \multicolumn{4}{l}{Model Expression in {\sc xspec} v12.8.2}\\
{\tt Comptonisation} & \multicolumn{4}{l}{{\tt CONSTANT*WABS*ZWABS*( DISKBB+ NTHCOMP +
  COMPTT + KDBLUR*PEXMON )}} \\
\hline
Spectrum & Component & Parameter & Value & Value\\
             &&& \multicolumn{1}{l}{Obs-1} & \multicolumn{1}{l}{Obs-2}\\
\hline
{\tt Mean Spec}   & {\tt ZWABS}     & N$_{H}$ (10$^{22}$ $cm^{-2}$) & 0~$^{+0.82}_{-0}$  & 0.027~$^{+0.014}_{-0.015}$\\
	     & {\tt DISKBB}        & T$_{in}$ (keV)        & 0.11~fixed						 & 0.11~fixed\\
             & {\tt DISKBB}        & norm                       & 135~$^{+59}_{-53}$				 & 135~fixed\\
             & {\tt NTHCOMP}   & $\Gamma$             & 2.26 fixed			 			 & 2.26 fixed\\
             & {\tt NTHCOMP}   & kT$_{seed}$ (keV) & tied to T$_{in}$		 			 & tied to T$_{in}$\\
             & {\tt NTHCOMP}   & kT$_{e}$ (keV)       & 100 fixed						 & 100 fixed\\
             & {\tt NTHCOMP}   & norm                       & (1.26~$^{+0.10}_{-0.11}$)$\times10^{-4}$  & (8.11~$^{+0.57}_{-0.80}$)$\times10^{-5}$\\
             & {\tt COMPTT}      & kT$_{seed}$ (keV) & tied to T$_{in}$		 			 & tied to T$_{in}$\\
             & {\tt COMPTT}      & kT (keV)                  & 0.38~$^{+0.61}_{-0.12}$			 & 0.21~$^{+1.14}_{-0.07}$\\
             & {\tt COMPTT}      & $\tau$                     & 10.9~$^{+6.6}_{-5.8}$			         & 19.2~$^{+15.4}_{-17.2}$\\
             & {\tt COMPTT}      & norm                       & (2.68~$^{+1.05}_{-1.66}$)$\times10^{-3}$  & (1.81~$^{+1.30}_{-1.68}$)$\times10^{-3}$\\
             & {\tt KDBLUR}      & R$_{in}$ ($R_{g}$)  & 20 fixed						& 20 fixed\\
             & {\tt PEXMON}     & Rel$_{refl}$             & -1.0 pegged					& -1.0 pegged\\
{\tt Cov. Spec} & {\tt NTHCOMP}   & norm            & (5.86~$^{+1.41}_{-1.73}$)$\times10^{-5}$  & (2.21~$^{+1.61}_{-1.67}$)$\times10^{-5}$\\
	    & {\tt COMPTT}    & norm                         & (1.45~$^{+4.77}_{-1.43}$)$\times10^{-4}$  & (9.74~$^{+12.81}_{-9.74}$)$\times10^{-4}$\\
	    &				&$\chi^{2}_{\nu}$        & $298/314=0.95$                                     & $176/224=0.78$ \\
\hline
{\tt Reflection} & \multicolumn{4}{l}{{\tt CONSTANT*WABS*ZWABS*( DISKBB + NTHCOMP +
  KDBLUR*RFXCONV*NTHCOMP )}} \\
\hline
Spectrum & Component & Parameter & Value & Value\\
             &&& \multicolumn{1}{l}{Obs-1} & \multicolumn{1}{l}{Obs-2}\\
\hline
{\tt Mean Spec}   & {\tt ZWABS}     & N$_{H}$ (10$^{22}$ $cm^{-2}$) & 0~$^{+0.02}_{-0}$ & 0.006~$^{+0.031}_{-0.005}$\\
	     & {\tt DISKBB}     & T$_{in}$ (keV)                   & 0.11~fixed 	& 0.11~fixed\\
             & {\tt DISKBB}     & norm                       	& 211~$^{+38}_{-136}$ & 83~$^{+117}_{-86}\times 10^{-4}$\\
             & {\tt NTHCOMP}   & $\Gamma$           & 2.42~$^{+0.07}_{-0.11}$       	& 2.22~$^{+0.10}_{-0.12}$\\
             & {\tt NTHCOMP}   & norm                    & 5.72~$^{+3.80}_{-4.00}\times 10^{-5}$& 3.60~$^{+3.00}_{-2.77}\times 10^{-5}$\\
             & {\tt KDBLUR}    & Index                      	& 5~fixed				& 5~fixed\\
             & {\tt KDBLUR}    & R$_{in}$ ($R_{g}$) & 4.98~$^{+4.30}_{-2.53}$		& 3.05~$^{+1.30}_{-3.05}$\\
             & {\tt RFXCONV}   & Rel$_{refl}$          & -2.19~$^{+1.14}_{-\infty}$		& -2.29~$^{+1.61}_{-\infty}$\\
             & {\tt RFXCONV}   & Fe$_{abund}$      & 1.39~$^{+2.50}_{-0.71}$		& 0.81~$^{+0.84}_{-0.81}$\\
             & {\tt RFXCONV}   & log~$\xi$                & 3.37~$^{+0.37}_{-0.22}$		& 2.86~$^{+0.30}_{-0.45}$\\
{\tt Cov. Spec} & {\tt NTHCOMP}   & norm  & 4.90~$^{+3.53}_{-3.57}\times 10^{-5}$& 1.08~$^{+1.97}_{-0.63}\times 10^{-5}$\\
	    & {\tt RFXCONV}   & Rel$_{refl}$          & -0.28~$^{+\infty}_{-\infty}$		& -2.60~$^{+2.40}_{-\infty}$\\
	    &&$\chi^{2}_{\nu}$& $295/312=0.95$ & $170/222=0.77$ \\
\hline
     \end{tabular}
 \label{tab-specfit}
 \end{minipage}
\end{table*}

To perform spectral fitting, we first applied the spectral model to
the entire 0.3-10 keV spectrum, but we found some parameters were
not well constrained because of parameter degeneracy. Therefore we
attempt to constrain the hard X-ray Comptonisation and neutral
reflection components by applying them only to the 2-10 keV spectra of
Obs-1 and Obs-2. The model fits the 2-10 keV spectra
well ($\chi^{2}_{v}=84/82$). The hard X-ray photon index
($\Gamma$) is found to be 2.26$\pm$0.12, the inner radius parameter
(R$_{in}$) of ${\tt kdblur}$ is 20$^{+380}_{-18}$ with reflection
solid angle $\Omega/2\pi$ fixed at unity. Removing the neutral
reflection component leads to $\Gamma=2.13\pm0.12$, but the fitting statistics
is only slightly worse ($\chi^{2}_{v}=96/84$). The low signal-to-noise in the
hard X-ray band cannot constrain this reflection component well.
Then we re-fit the model to the entire 0.3-10
keV band by fixing $\Gamma$ at 2.26 and R$_{in}$ at 20, and all
spectral components are better constrained then before.
The best fit parameters are listed in Table~\ref{tab-specfit}. 

In Obs-1, the HF covariance spectrum can be well fitted by the hard
Comptonisation with $\Gamma=2.26$, without significant contribution
from a variable soft Comptonisation component. On the contrary, the HF
covariance spectrum in Obs-2 is softer so requires more contribution from
the soft Comptonisation (Figure~\ref{fig-specfit}a and Figure~\ref{fig-specfit}b).
Neither covariance spectrum requires any disc component.

We can also test for the presence of an accretion disc component in
the soft X-ray bandpass by removing the disc component from the model
and then re-fitting the Obs-1 spectra. The electron temperature (kT)
of the soft Comptonisation component increases from 0.38 keV to 0.46
keV, the optical depth ($\tau$) decreases from 10.9 to 7.1, the
normalisation increases by 10\%. However, $\chi^{2}_{v}$ increases
from 0.94 to 0.98 (i.e. $\Delta\chi^2=10$ for the removal of 2 free
parameters, which is only marginally significant). The significance
increases if we consider only the 0.3-0.5 keV band, where
$\chi^{2}_{v}$ increases from 1.22 to 1.33 after removing the disc
component. Thus both the time average spectra and HF variability
support the decomposition of the soft X-ray spectrum into at least two
components, which we interpret here as 
a contribution from the accretion disc, and an additional  low temperature
Comptonisation. A high temperature Compton spectrum
is required to reproduce the 2-10~keV spectrum.

\begin{figure*}
\begin{tabular}{cc}
\includegraphics[scale=0.45]{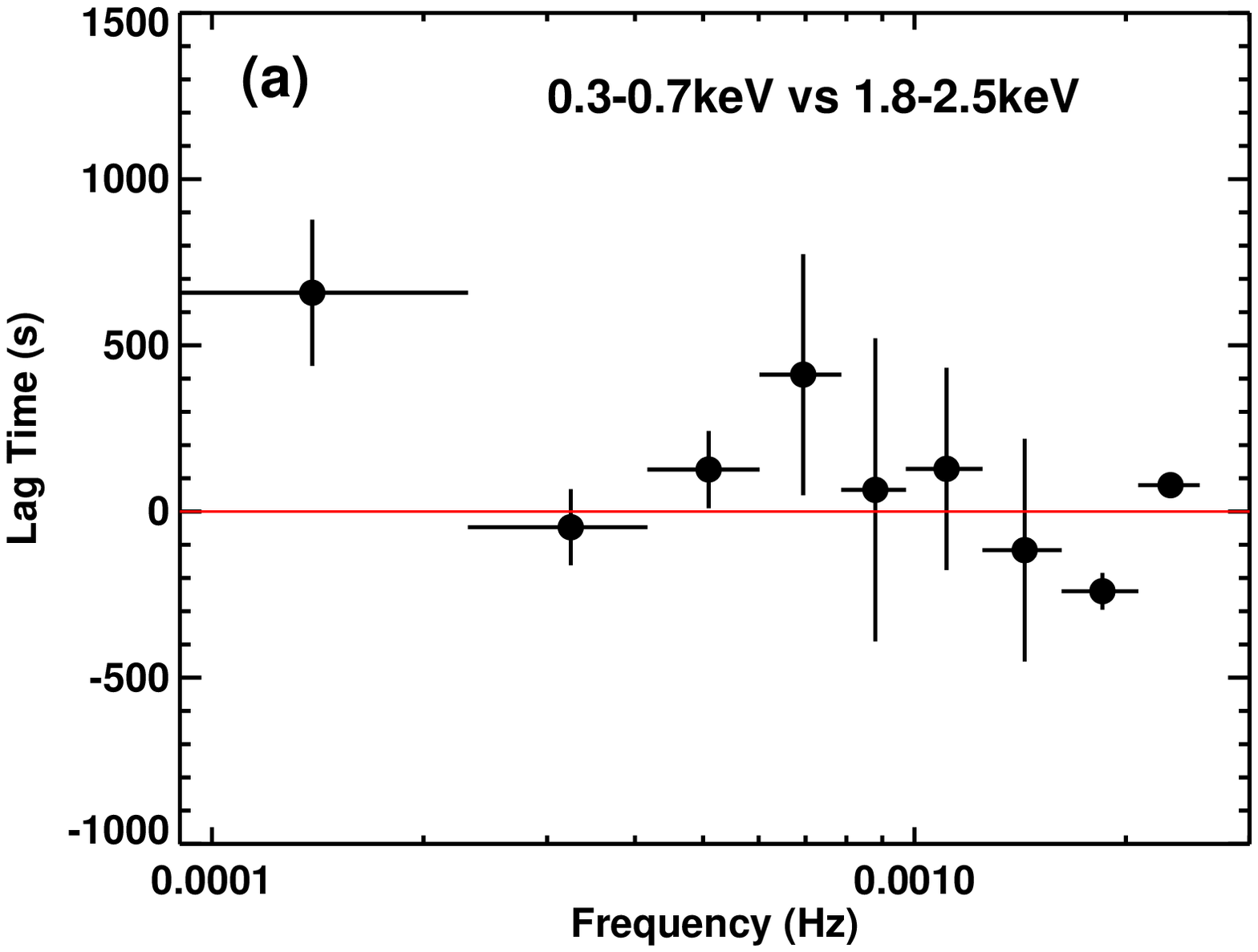}  & \includegraphics[scale=0.45]{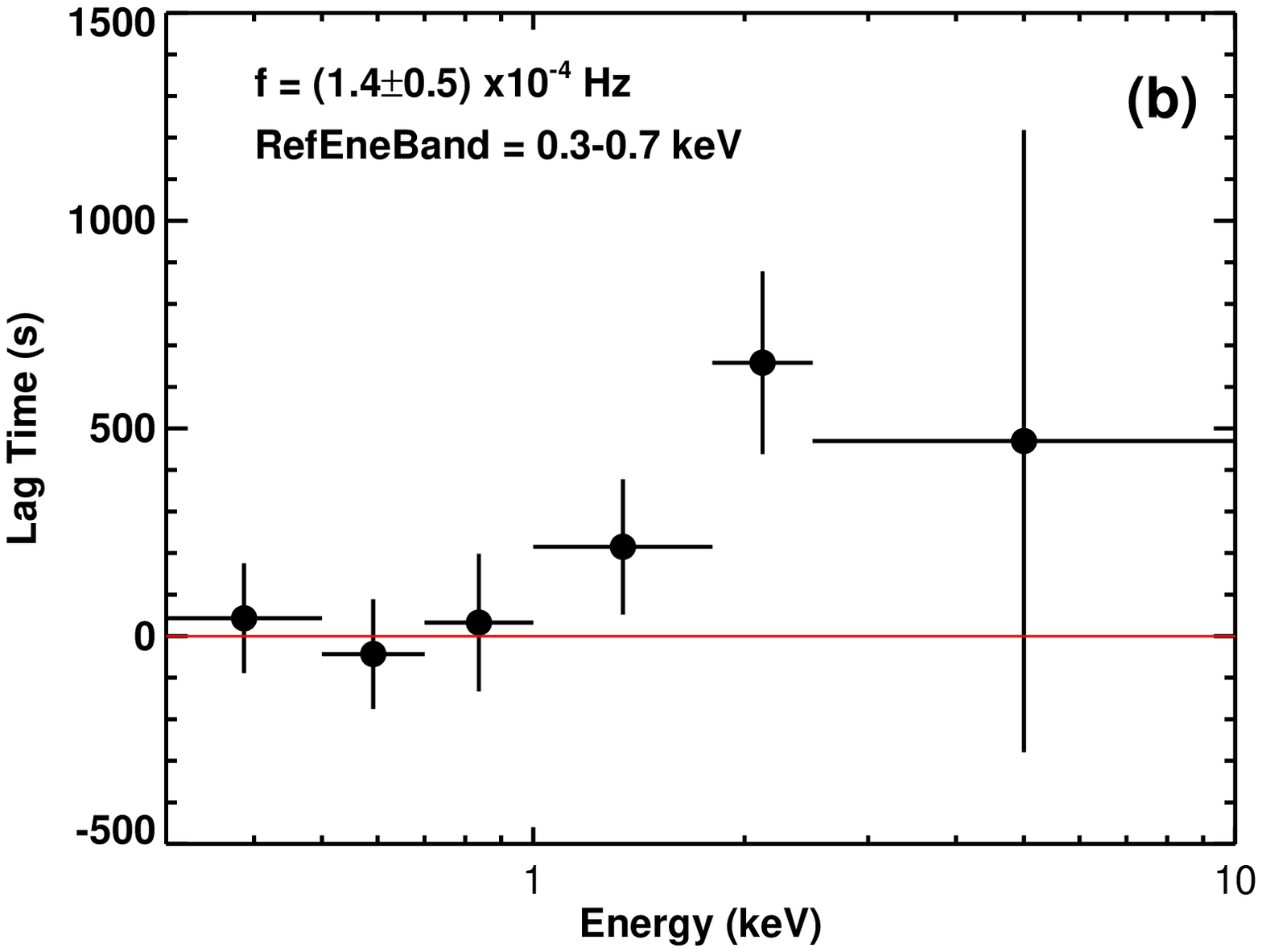} \\
\includegraphics[scale=0.45]{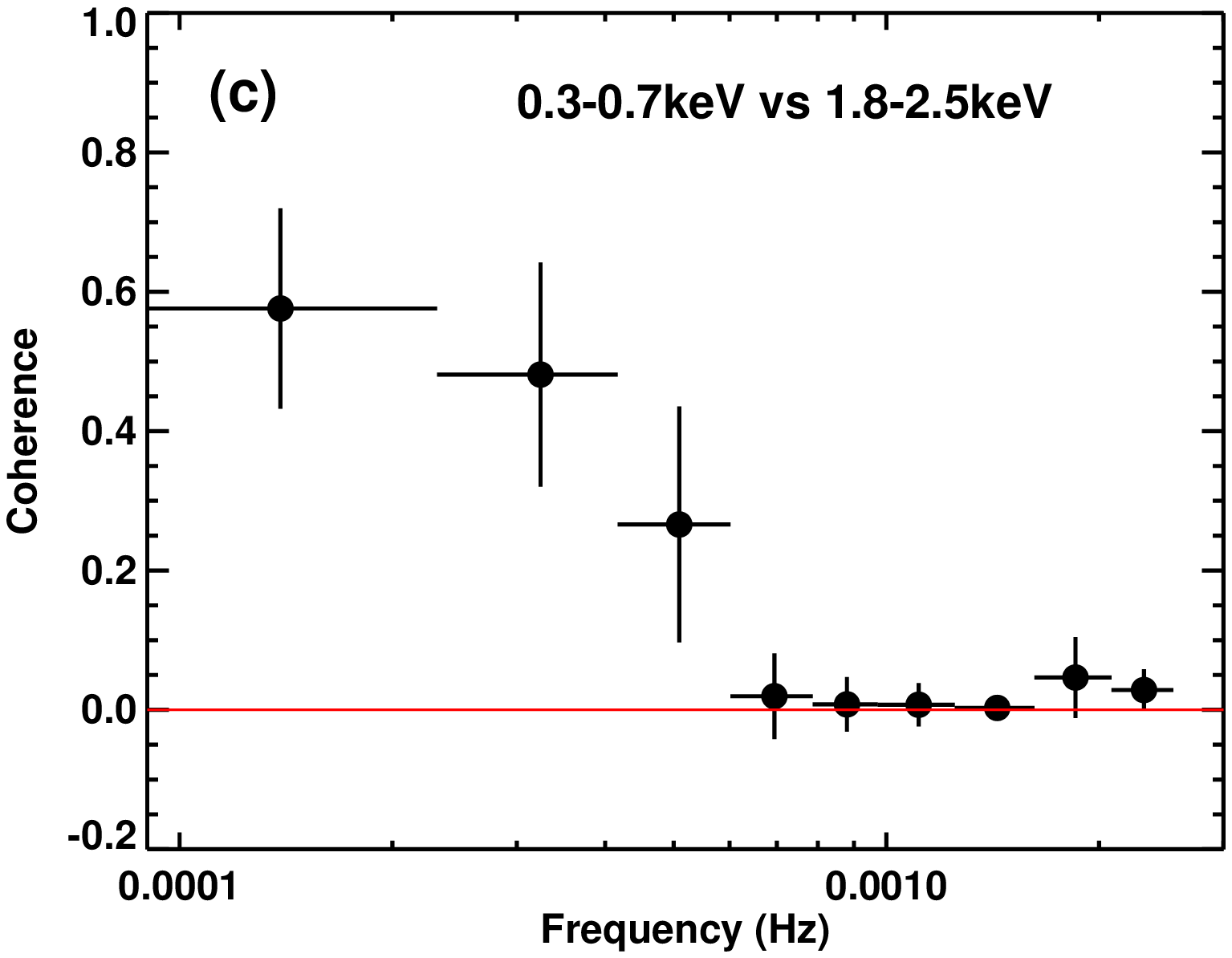} & \includegraphics[scale=0.45]{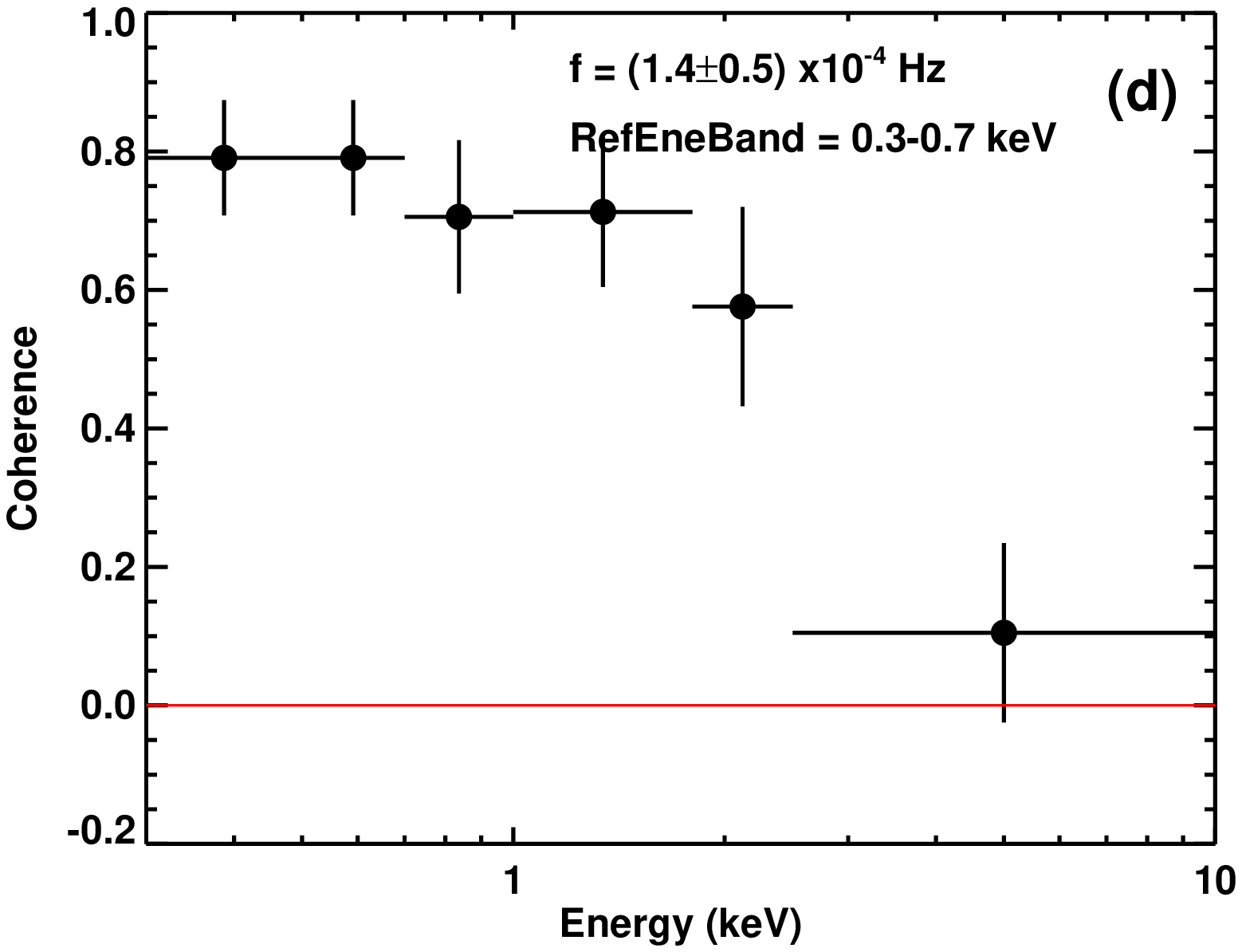} \\
\end{tabular}
\caption{Frequency and energy resolved time-lag and coherence for RX1140. The points are derived from five segments of 11 ks length each,
  which comprise three segments in Obs-1 and two segments in Obs-3 (see
  Section~\ref{sec-var}).}
\label{fig-lag}
\end{figure*}

\subsection{Reflection Model}
\label{sec-refl}

Ionised, strongly smeared disc reflection plus a thermal accretion
disc component was previously reported to be a good model for the
time-averaged Obs-1 spectrum of RX1140 (Miniutti et al. 2009, Ai et al.
2011). Thus we use {\tt diskbb+nthcomp+kdblur*nthcomp} as our spectral
model, with T$_{in}$ of the disc component fixed at 110 eV as
before. Figure~\ref{fig-specfit}c and Figure~\ref{fig-specfit}d
shows this model applied
to Obs-1 and Obs-2 respectively. Again, the low statistics mean that
some parameters cannot be well constrained. We choose to fix the
illumination index of the smeared reflection at $5$, and fit for the
inner disc radius. This is always small in this model, at $\sim
3-5R_g$, so any reverberation timescale should be short.  Hence both
the high temperature Comptonisation and its reflection should vary
together. However, the relatively flat HF covariance spectrum in Obs-1
favours a less reflection component contribution than in the time averaged spectrum,
while the softer HF covariance spectrum in Obs-2 favours a larger reflection contribution.
But again the poor signal-to-noise means
that this is not very significant (Table~\ref{tab-specfit}).
Similar as in the Comptonisation scenario,
accretion disc component is not detected in the HF covariance spectra.

\subsection{Lag and coherence spectra}

Both Comptonisation and reflection models for the soft X-ray excess
can give comparably good fits to both the time averaged and HF
covariance spectra. Both models also require that there is an
additional soft component, which we interpret as the direct disc
emission, at the softest energies. 

We can also use spectral coherence and time lags to constrain the X-ray
spectral mechanisms (e.g. M$^{c}$Hardy et al. 2004; Fabian et al.
2009). However, the low count rate of RX1140 and the
relatively short continuously sampled light curve segments ($\la$11
ks, see Section~\ref{sec-var}) mean that the statistics are
limited. We calculate these as in Nowak et al. (1999), 
using only Obs-1 and 3 (since Obs-2 has a significantly different
spectrum and variability). We choose low and high energy band
with 0.3-0.7~keV and 1.8-2.5~keV so as to try to best isolate
different spectral components whilst still having enough counts. 

Figure~\ref{fig-lag}a shows that there is a soft reverberation lag of
$\sim 200$~s seen between the 0.3-0.7~keV and 1.8-2.5~keV light curves
at a frequency of $\sim 2\times 10^{-3}$~Hz. But the zero spectral
coherence and the dominance of Poisson noise
at this frequency means this lag can be artificial.
However, there is a clear
positive lag at the lowest frequencies (below $2\times 10^{-4}$~Hz),
where the soft leads the harder band.
This lag is supported by the good spectral coherence between
these bands up to $6\times 10^{-4}$~Hz
(Figure ~\ref{fig-lag}c and Figure ~\ref{fig-lag}d).
Note that Poisson noise will not be an issue on such long timescales.
Figure~\ref{fig-lag}b) shows
the energy spectrum of this long timescale soft lead, showing that
there is a clear break between the time lags at low and high
energy, although we do not have enough statistics at hard X-ray.
This is not easily compatible with the reflection model shown
in Figure~\ref{fig-specfit}b), where a single component dominates
the spectrum from 0.4-10~keV.

Thus while the signal-to-noise is poor, these data provide marginal evidences
in terms of spectral coherence and time-lag to support
a model where the majority of the soft X-ray excess does not 
contribute to the 2-10~keV bandpass, favouring the low temperature
Comptonisation origin over the reflection dominated model.

\section{Broad band Spectral Energy Distribution}
\label{sec-sed}

\subsection{Multi-wavelength Data}

As outlined in the introduction, RX1140 was reported as an
IMBH, whose mass is estimated to be $5-10\times 10^{5}M_{\odot}$
accreting at a high fraction of the Eddington limit.  We construct a
multi-wavelength SED based on non-simultaneous observations
(Figure~\ref{fig-sed1}). We use the {\it XMM-Newton} PN spectrum from
Obs-1 taken in 2005, together with archival data from the {\it ROSAT}
PSPCB (1992-06-16) which we reduced using {\tt XSELECT} v2.4b.  The
spectra were regrouped fixing a minimum of 25 counts per bin, and are
in good agreement with the time-averaged PN {\it XMM-Newton} spectrum.

In the optical and UV band, there are simultaneous photometric points
from {\it XMM-Newton} OM for Obs-1 (UVM2 and UVW1 filters) and Obs-2
and Obs-3 (UVW1, U and B filters).  The UVW1 filter fluxes of Obs-1 and
Obs-3 are very similar, so are their X-ray fluxes, so we use all OM
data from these two observations (except for UVW1 since its waveband
is already covered by the combination of UVM2, U and B filter bands).  We do
not use Obs-2 as it is clearly different in optical/UV as well as
X-ray, with only 90\% flux in UVW1, 84\% in U and 99\% in B compared
to Obs-3. These are less suppressed than the X-ray flux (Obs2/Obs3 =
46\% in 0.3-2 keV, 63\% in 2-10 keV), indicating that the amplitude of
UV/optical variability is less than that of the X-ray over variability
timescales of weeks and years.

Other data points in Figure~\ref{fig-sed1} include {\it GALEX}
photometry (obs-date: 2007-03-22), SDSS {\sc ugriz} photometry
and fibre spectrum (obs-date: 2001-03-26), {\it
UKIDSS} Petrosian magnitude of JYHK photometry (obs-date: 2008-04-21).
The difference in the normalisation, especially that for the {\it
UKIDSS} and SDSS fibre fluxes, is likely due to different aperture
sizes.

\subsection{Broadband SED Fitting}
\label{sec-sed-fit}

We use {\tt optxagnf} in {\sc xspec} v12.8.2, to fit the SED from the
optical to hard X-ray energies (D12).  This includes both
intrinsic disc emission, together with low and high temperature
Comptonisation, as favoured by the spectral fits in Section~\ref{sec-specfit}. 
An additional
neutral reflection ({\tt kdblur*pexmon} model, Laor 1991, recoded as
a convolution model by Nandra et al. 2007) is added to fit the
marginally rising shape above 4 keV. We fix the inclination of the
reflector at $30^\circ$, and fix the spectral index of the hard
Comptonisation component at 2.26 (see Section~\ref{sec-specfit}).  We
correct for extinction by a Galactic column density of
$N_{H}~=~1.91{\times}10^{20}~cm^{-2}$ (Kalberla et al. 2005). We also
allow for intrinsic extinction in the host galaxy. This gives a best
fit around $N_{H}~=~2{\times}10^{20}~cm^{-2}$ at z = 0.0811 ({\tt
zwabs} model, Morrison \& McCammon 1983). Both Galactic and intrinsic
reddening are assumed to follow the standard dust to gas conversion
formula of $E(B-V)~=~1.7{\times}10^{-22}N_{H}$ (Bessell 1991) and
modelled by {\tt redden} and {\tt zredden} model (Cardelli et al. 1989)
in {\sc xspec}.

\begin{figure}
\includegraphics[scale=0.45]{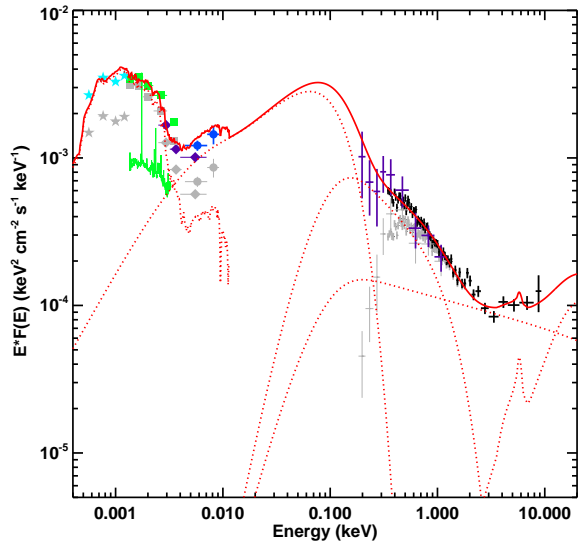}
\caption{Broadband SED of RX1140.  Black data in the X-ray
band is the Obs-1 mean spectrum, purple data in the soft X-ray band is
from {\it ROSAT}. The UV, optical, infrared photometric points
comprises {\it GALEX} (blue circles), OM of {\it XMM-Newton} (purple
diamonds), SDSS {\sc ugriz} (green squares) and fibre (green
spectrum), {\it UKIDSS} JYHK (cyan stars). The best fit {\tt optxagnf}
model assuming black hole mass as a free parameter is shown as a red
solid line is the best-fit model, orange dotted lines show different
SED components. This gives $M\sim 10^{7}M_\odot$, much higher than
expected from H$\beta$ line width and reverberation mapping.}
\label{fig-sed1}
\end{figure}

The far UV {\it GALEX} points fit nicely on the expected disc spectrum
from {\sc optxagnf}, but the optical and near UV points display a
curvature similar to that of host galaxy emission. The SDSS images
show that the host galaxy of RX1140 is of an Sc type galaxy.
Therefore we select an Sc galaxy template from the SWIRE library
(Polletta et al. 2007) and incorporate this into our {\sc xspec}
model.  A constant normalisation offset is allowed between the {\it
UKIDSS} points and other data.

Figure~\ref{fig-sed1} shows the extinction/reddening/aperture
corrected data (coloured points, compared to the observed data points
in grey) and the best-fit SED model.  The formal fitting statistics
are poor (reduced $\chi^{2}~=~982/330$), which is mainly due to the
small error bars of the photometric points in the optical/UV band.

\begin{figure*}
\begin{tabular}{cc}
\includegraphics[bb=90 216 576 684, scale=0.45]{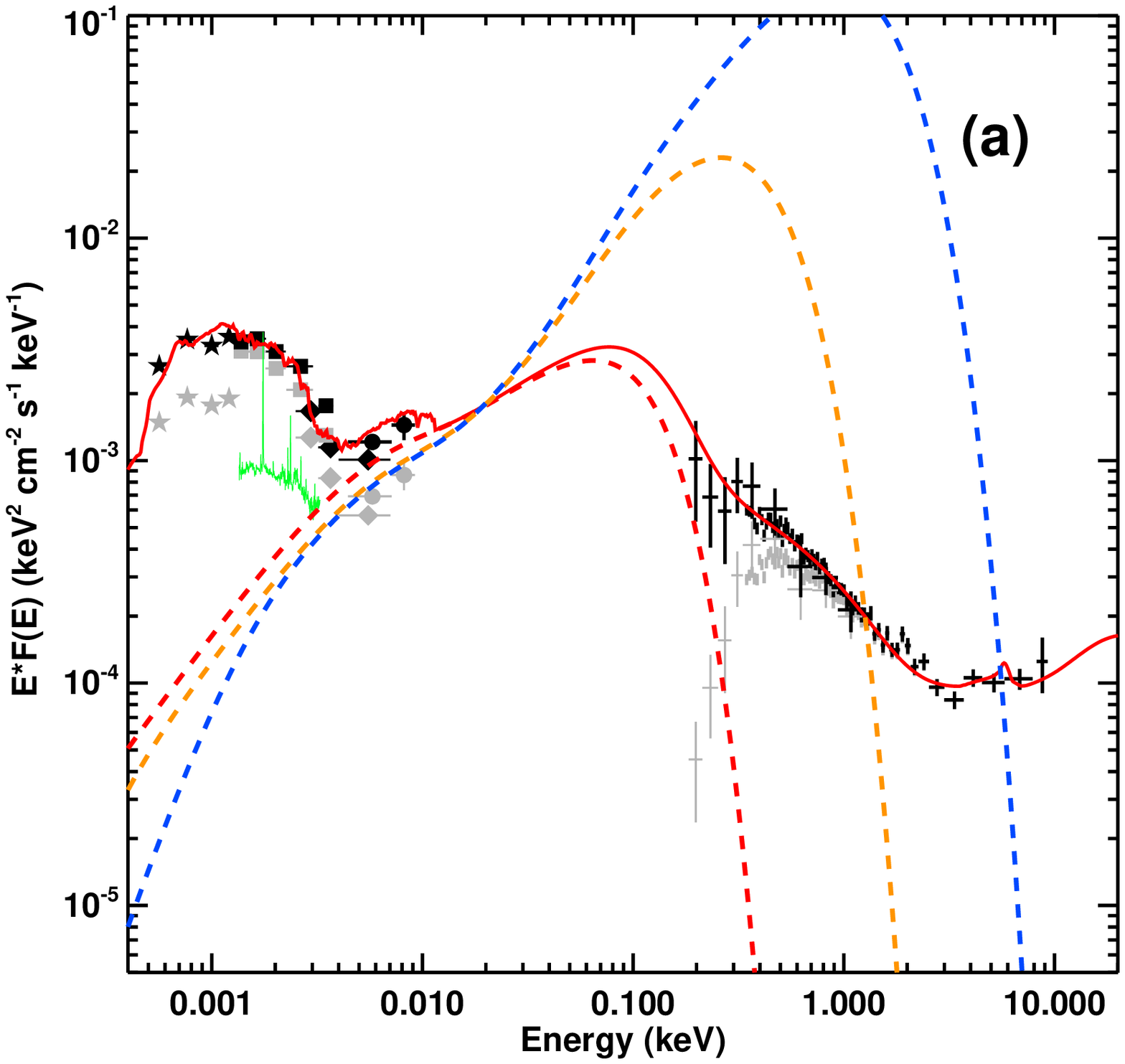} &
\includegraphics[bb=30 216 576 684, scale=0.45]{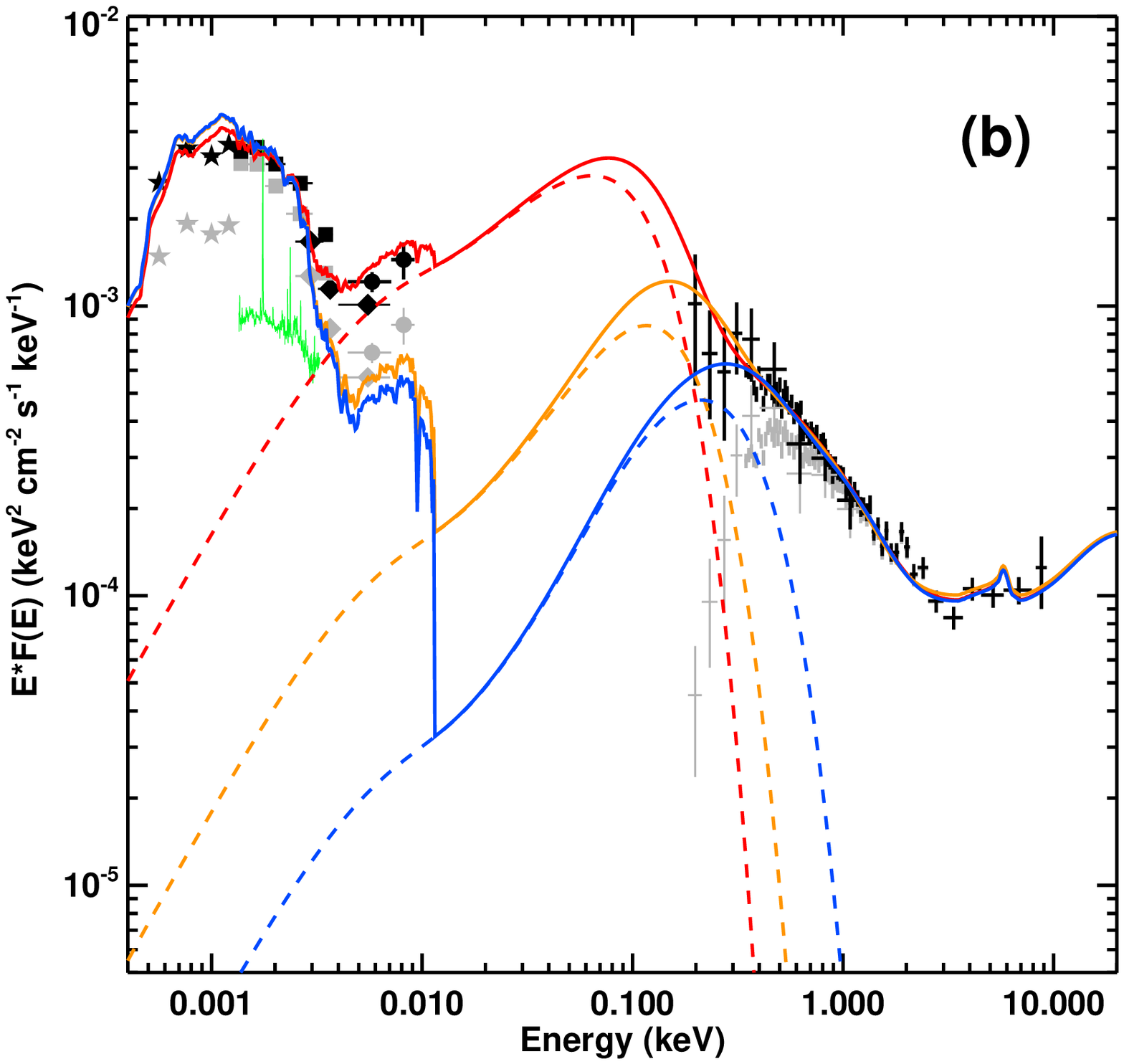} \\
\end{tabular}
\caption{Broadband SED fitting similar as in Figure~\ref{fig-sed1},
  but with different black hole masses: $M=9.6\times10^{6}M_{\odot}$
  (red), $1.0\times10^{6}M_{\odot}$ (blue), $1.5\times10^{5}M_{\odot}$
  (green). In Panel-a, the models primarily fit the UV points and have
  $L/L_{Edd}=0.17, 10$ and $400$, respectively for disc emission
  outside 15 $R_{g}$ (see Section~\ref{sec-sed-fit}). In Panel-b, the
  models tend to primarily fit the X-ray spectra and have
  $L/L_{Edd}=0.17,0.56$ and $2$, respectively. Solid lines are the
  total models, while dashed lines are the disc components.}
\label{fig-sed2}
\end{figure*}

However, the best-fit black hole mass is $9.6\times10^{6}M_{\odot}$
(with correspondingly fairly low $L/L_{Edd}=0.17$). This is almost an
order of magnitude higher than the expected mass from optical
spectroscopy and reverberation (Greene, Ho \& Barth 2008; Rafter et
al. 2011).  The Eddington ratio is correspondingly low,
$L/L_{Edd}=0.17$, which is in conflict with the classification of this
object as a NLS1, as supported by the steep X-ray spectrum, and by the
{\sc optxagnf} coronal radius of $\sim 15 R_{g}$, consistent with the
typical value for NLS1 (J12c, D12).

We test how robustly the SED modelling constrains mass by fitting pure
disc models down to $15R_g$ (dashed lines) to the UV points for masses of
$1.5\times10^{5}$ (blue: $L/L_{Edd}=400$), $1.0\times10^{6}~M_{\odot}$
(orange: $L/L_{Edd}=10$) and $1.0\times10^{7}~M_{\odot}$ (red:
$L/L_{Edd}=0.17$ as before. The red solid line shows the previous best
fit. These disc component all give optical fluxes at $5100${\AA} which
match to the observed variable flux (Rafter et al. 2011). However, all
the reasonable masses over-predict the observed X-ray emission by
factors of 10-100. Allowing the disc to extend down to $6R_g$, or
$1.2R_g$ for a maximally spinning black hole, would
make this discrepancy worse (e.g. Done et al. 2013). 

In the right panel of Figure~\ref{fig-sed2}, we use the same range of
masses but aim instead to primarily fit the X-ray spectrum, and
extrapolate the model to lower energies. The model with a
$10^{7}~M_{\odot}$ (red curve) can fit the far UV (and variable
optical) data, whereas the model for the $10^{6}$ and $1.5\times
10^5~M_{\odot}$ black hole masses under-predict the disc continuum by a
factor of 10-50 as the peak of the disc spectrum shifts into the soft
X-ray band, and its contribution to the UV and optical decreases.

\section{Discussion} 
\label{sec-discussion}

\subsection{Black Hole Mass}
\label{sec-mass}

Section~\ref{sec-introduction} summarised previous mass estimates for
RX1140 from RM (Rafter et al. 2011) and H$\beta$.
All previous black hole masses based on optical
Spectra report RX1140 to be an IMBH with $M\la
10^{6}M_{\odot}$.

\begin{figure}
\includegraphics[bb=90 216 540 576, scale=0.55]{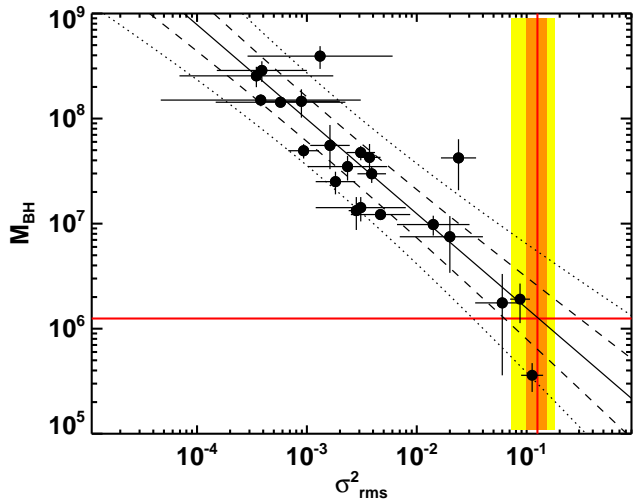}
\caption{M-$\sigma^{2}_{rms}$ relation formed by the RM sample, where
  $\sigma^{2}_{rms}$ is the excess variance of 2-10 keV for 40 ks
  timescale. $\sigma^{2}_{rms}$ values are taken from Ponti et al. (2012), the
  RM masses are taken from Peterson et al. (2004, 2005), Denney et al. (2006),
  Du et al. (2014). The black solid line is the regression line assuming
  $\sigma^{2}_{rms}$ as an independent variable. Dashed and dotted lines show
  the $\pm1\sigma$ and $\pm2\sigma$ confidential ranges for new observation. Red
  vertical line indicates the mean $\sigma^{2}_{rms}$ for RX1140, with
  $\pm1\sigma$ intervals determined from Obs-1 (orange) and Obs-2 (green). The
  horizontal red line indicates the mass of RX1140 predicted by the
  regression line.}
\label{fig-massrms}
\end{figure}

The X-ray variability power spectra also strongly support such a low mass,
with no observed high frequency break on timescales longer than 500~s
(see Figure.~\ref{fig-powspec}). This means that the total excess
variance, $\sigma_{rms}^2$, is high. We include our new data on the
$M-\sigma_{rms}^{2}$ relation (Miniutti et al. 2009; Zhou et al. 2010;
Ponti et al. 2012). Figure~\ref{fig-massrms} shows the reverberation
mapped subsample of objects from Ponti et al. (2012), where
$\sigma_{rms}^{2}$ values were calculated from 40 ks light curves in
the 2-10~keV bandpass. The best fit linear regression line is also
shown. For RX1140, $\sigma_{rms}=0.11\pm0.04$ in Obs-1 and
$0.12\pm0.08$ in Obs-2 (Obs-3 does not have sufficiently good data). These
values correspond to $M=1.25{\times}10^{6}M_{\odot}$. A
$10^{7}M_{\odot}$ mass lies outside the 2$\sigma$ region, but any
mass within $3{\times}10^{5}-3.0{\times}10^{6}M_{\odot}$ is 
consistent with the relation (see also Ludlam et al. 2015). 

This highlights the inconsistency between the black hole mass derived
from the SED models of $\sim 10^7M_\odot$, and all other mass
determination methods which give $10^6M_\odot$. For a mass of
$10^6M_\odot$, the SED models can either fit the optical/UV flux from
the AGN, but then over-predict the X-rays, or they can fit the X-rays, but
under-predict the optical/UV (Figure~\ref{fig-sed2}b).  Given that the
fraction of optical/UV flux from the AGN is constrained by the observed variability
(Rafter et al. 2011), then this requires that the 
$10^6M_\odot$ disc fit in Figure~\ref{fig-sed2}a) correctly
describes the outer disc, with a derived $L/L_{Edd}=10$. Clearly this is
substantially super-Eddington, so there can be substantial energy loss
via optically thick advection and/or winds from the inner disc. The
fitted SED models, which assume that the disc is sub-Eddington so that
neither of these processes is important
(orange and blue curves in Figure~{fig-sed2}a),
then over-estimate the total luminosity.

The extent of the energy losses due to winds/advection can 
be estimated from the best fit SED model. The parameters of 
$M=10^7M_\odot, L/L_{Edd}=0.17$ give $L_{bol}\sim 2\times
10^{44}$~ergs~cm$^{-2}$~s$^{-1}$, which is around the Eddington limit for the
most probable mass of $10^6M_\odot$. This is a supporting evidence for the
proposal of Wang et al. (2014) that super-Eddington black holes have
total luminosity which saturates around the Eddington limit.

However, RX1140 also has another unusual feature in that it exhibits
strong optical variability, as seen in the RM campaign of Rafter et al.
(2011). The majority of NLS1 show less optical variability than the broader
line Seyferts (e.g. Ai et al. 2013, but see Kelly et al. 2013),
which is somewhat surprising in view of
their higher X-ray variability (e.g. Ponti et al. 2012). This
unusually strong optical variability could indicate that some of the
X-rays are reprocessed within a large scale wind. In which case the models above in
Figure~\ref{fig-sed2}a, which require only disc emission to fit the
optical, may be overestimating the mass accretion rate.  Nonetheless
there must be sufficient X-ray emission to power an additional optical
component. The X-ray model fit based on $10^6M_\odot$ shown in
Figure~\ref{fig-sed2}b has an X-ray peak which has similar ${\nu}L_{\nu}$
to the required far UV emission, so this model with $L/L_{Edd}=0.56$
would require that almost all the X-rays are reprocessed. A more
likely reprocessing fraction of 10-50\% then requires a higher
Eddington ratio of $1-5$. So perhaps the most plausible model
is a combination of the two $10^6M_\odot$ models in
Figure~\ref{fig-sed1}a and Figure~\ref{fig-sed1}b,
where the X-ray emission is partly suppressed
by wind/advection losses and the optical emission is correspondingly
enhanced by reprocessing in the wind.

We note that if winds and advection do play a significant role, it is curious 
that the X-ray spectrum and variability properties are not more
noticeably different from other well studied NLS1 such as PG 1244+026, RE J1034+396
and RX J0136.9-3510 (e.g. Middleton et al. 2009; Jin et al. 2009; Jin
et al. 2013). This then raises the possibility that these sources are
likewise super-Eddington.  PG~1244+026 has an SED which is acceptably 
well fit by a $10^7M_\odot$ black hole accreting at about $L_{Edd}$
(Done et al. 2013). This SED has a distinct outer disc spectral shape in
the optical/UV (e.g. Done et al. 2013), which constrains
$(M\dot{M})^{2/3}\propto (M^2 L/L_{Edd} )^{2/3}$. Reducing the mass by
a factor of 4 to $2.5\times 10^6M_\odot$ i.e. removing the Marconi et
al. (2008) radiation pressure correction to the H$\beta$ line width,
then requires $L/L_{Edd}=16$, which clearly means that this
source could be very super-Eddington.
However, this is probably an over-estimate as
the well defined HF break in the power spectrum of this source
(Jin et al. 2013) suggests at least a factor 5 higher mass than in
RX1140, but this still gives $L/L_{Edd}=4$. Advection and winds could
also be significant in this source, which would invalidate the black
hole spin determination described in Done et al. (2013).

If the majority of NLS1 are super-Eddington, but the typical Broad
Line Seyfert 1's have $L/L_{Edd}\sim 0.05-0.2$,
then this implies a distinct bimodal (or a very long tail), mass accretion rate distribution.
Also, such large energy losses due to advection/winds are probably
not seen in the majority of Ultra-luminous X-ray sources (e.g. Sutton
et al. 2015), although they may be present in the more extreme
sources discussed by Gladstone, Done \& Roberts (2009). The current data suggest that most of
these sources represent moderately super-Eddington flows onto stellar mass
black holes with masses around $30M_\odot$ compared with those with $\sim
10M_\odot$ seen in our Galaxy, due to their lower metallicity resulting
in less mass loss from winds: Zampieri \& Roberts (2009). Theoretical
models predict quite high efficiency for these flows (Jiang, Stone \& Davis 2014,
but see also S{\c a}dowski et al. 2015 who report lower
efficiency). Clearly there remain are many aspects of super-Eddington
accretion flows which are not yet understood.

\section{Summary}
\label{sec-summary}

In this paper we use new {\it XMM-Newton} data from one of the lowest
mass AGN known to try to understand the nature of the soft X-ray
excess, and how this fits in the context of its broad band
SED. The mass from H$\beta$ (uncorrected for radiation pressure:
Marconi et al. 2008), upper limit from reverberation mapping and fast
X-ray variability are all consistent with a mass of
$10^6M_\odot$. This, together with a steep 2-10~keV X-ray spectrum,
and even steeper soft X-ray emission in this object make this a clear
member of the `simple' NLS1 class of AGN, where all objects with low black
hole mass accreting at a high accretion rate.

We fit the time averaged X-ray data together with the covariance
spectra of the fast variability to show that both low temperature
Comptonisation and highly smeared, ionised reflection models can fit
the data, but that both require an additional component at the lowest
energies which we interpret as being from the inner disc.  However,
the {\it RMS}, lag and coherence spectra all point to a break in 
variability properties between the soft and hard X-ray bands, which is
not consistent with the reflection models.  Instead these support the
low temperature Comptonisation model as this does not extend into the
2-10~keV band. All these properties are similar to other `simple' NLS1
such as PG~1244+026 (Jin et al. 2013), RE J1934+396 (Middleton et al.
2009) and RX J0136.9-3510 (Jin et al. 2009).

However, putting the X-ray data into the context of the broadband SED
reveals something quite new. The far UV and variable optical flux
require an outer disc which has $L/L_{Edd}\sim 10$ for an assumed mass
of $10^6M_\odot$. This far over-predicts the soft X-ray emission for
standard efficiency even assuming zero spin for the black hole.  This
strongly requires substantial energy loss via advection and/or winds
unless the black hole mass is underestimated by a factor of 10, or the
variable optical flux is not predominantly from the outer accretion
disc (both of which seen extremely unlikely). If so, the other similar
NLS1 are also likely to be similarly super-Eddington. This removes the
requirement for low spin in these objects (Done et al. 2013), but also
makes it highly unlikely that the geometry of these NLS1 are well
represented by a flat disc, as used in reverberation studies.

\section*{Acknowledgements}
This work is based on observations obtained with {\it XMM-Newton}, an ESA
science mission with instruments and contributions directly funded by ESA Member
States and NASA. This work makes use of data from SDSS, whose funding is
provided by the Alfred P. Sloan Foundation, the Participating Institutions, the
National Science Foundation, the U.S. Department of Energy, the National
Aeronautics and Space Administration, the Japanese Monbukagakusho, the Max
Planck Society, and the Higher Education Funding Council for England. We have
made use of the {\it ROSAT} Data Archive of the Max-Planck-Institut f\"{u}r
extraterrestrische Physik (MPE) at Garching, Germany.


\begin{thebibliography}{}

\bibitem[Ai et al. (2011)]{2011ApJ...727..31A} Ai Y. L., Yuan W., Zhou H. Y., Wang T. G., Zhang S. H., 2011, ApJ, 727, 31

\bibitem[Ai et al.(2013)]{2013AJ....145...90A} Ai Y. L., Yuan W., Zhou H., Wang T. G., Dong X. B., Wang J. G., Lu H. L., 2013, AJ, 145, 90 

\bibitem[\protect\citeauthoryear{Arnaud}{1996}]{1996ASPC..101...17A} Arnaud K.~A., 1996, ASPC, 101, 17 

\bibitem[Barth et al. (2004)]{2004ApJ...607..90B} Barth A. J., Ho L. C., Rutledge R. E., Sargent W. L. W., 2004, ApJ, 607, 90

\bibitem[Baskin, Laor \& Stern (2014)]{2014MNRAS.438..604B} Baskin A., Laor A., Stern J., 2014, MNRAS, 438, 604 

\bibitem[\protect\citeauthoryear{Bentz et al.}{2006}]{Bentz06} Bentz M. C., et al., 2006, ApJ, 651, 775

\bibitem[\protect\citeauthoryear{Bentz et al.}{2009a}]{Bentz09a} Bentz M. C., Peterson B. M., Netzer H., Pogge R. W., Vestergaard M., 2009a, ApJ, 697, 160

\bibitem[\protect\citeauthoryear{Bentz et al.}{2009b}]{Bentz09b} Bentz M. C., et al., 2009b, ApJ, 705, 199

\bibitem[Bessell(1991)]{1991A&A...242L..17B} Bessell M. S., 1991, A\&A, 242, L17 

\bibitem[Boroson(2002)]{2002ApJ...565...78B} Boroson T. A., 2002, ApJ, 565, 78

\bibitem[\protect\citeauthoryear{Boroson \& Green}{1992}]{Boroson92} Boroson T. A., Green R. F., 1992, ApJS, 80, 109

\bibitem[Cardelli et al.(1989)]{1989ApJ...345..245C} Cardelli J. A., Clayton G. C., Mathis J. S., 1989, ApJ, 345, 245 

\bibitem[Davis \& Laor(2011)]{2011ApJ...728...98D} Davis S. W., Laor A., 2011, ApJ, 728, 98 

\bibitem[\protect\citeauthoryear{Denney et al.}{2006}]{Denney06} Denney K. D., et al., 2006, ApJ, 653, 152

\bibitem[\protect\citeauthoryear{Denney et al.}{2010}]{Denney10} Denney K. D., et al., 2010, ApJ, 721, 715

\bibitem[Done et al. (2012)]{2012MNRAS.420.1848D} Done C., Davis S. W., Jin C., Blaes O., Ward, M., 2012,  MNRAS, 420, 1848 (D12)

\bibitem[Done et al. (2013)]{2013MNRAS.434.1955D} Done C., Jin C., Middleton M., Ward M., 2013,  MNRAS, 434, 1955

\bibitem[Dong et al. (2007)]{2007ApJ...657..700D} Dong X., et al., 2007, ApJ, 657, 700

\bibitem[\protect\citeauthoryear{Du et al.}{2014}]{Du14} Du P., et al., 2014, ApJ, 782, 45

\bibitem[\protect\citeauthoryear{Edelson et al.}{2002}]{Edelson02}Edelson R., Turner T. J., Pounds K., Vaughan S., Markowitz A., Marshall H., Dobbie P., Warwick R., 2002, ApJ, 568, 610

\bibitem[Elvis et al.(1994)]{1994ApJS...95....1E} Elvis M., et al., 1994, ApJS, 95, 1

\bibitem[Fabian, Kara \& Parker(2014)]{2014efxu.conf..279F} Fabian A. C., Kara E., Parker M. L., 2014, Suzaku-MAXI 2014: Expanding the Frontiers of the X-ray Universe, 279 

\bibitem[Fabian \& Miniutti(2005)]{2005astro.ph..7409F} Fabian A. C., Miniutti, G., 2005, arXiv:astro-ph/0507409

\bibitem[Fabian et al.(2009)]{2009Natur.459..540F} Fabian A. C., et al., 2009, Nature, 459, 540

\bibitem[Fabian et al.(2013)]{2013MNRAS.429.2917F} Fabian A. C., et al., 2013, MNRAS, 429, 2917 

\bibitem[Filippenko \& Ho (2003)]{2003ApJ...588L..13F} Felippenko A. V., Ho L. C., 2003, ApJ, 588L, 13

\bibitem[Gardner \& Done(2015)]{2015MNRAS.448.2245G} Gardner E., Done C., 2015a, MNRAS, 448, 2245 

\bibitem[Gierli{\'n}ski \& Done(2004)]{2004MNRAS.349L...7G} Gierli{\'n}ski M., Done C., 2004, MNRAS, 349, L7

\bibitem[Gladstone et al.(2009)]{2009MNRAS.397.1836G} Gladstone J. C., Roberts T. P., Done C., 2009, MNRAS, 397, 1836 

\bibitem[\protect\citeauthoryear{Gierli\'{n}ski \& Done}{2006}]{Gierlinski06}Gierli\'{n}ski M., Done C., 2006, 
MNRAS, 371, L16

\bibitem[Greene \& Ho (2004)]{2004ApJ...610..722} Greene J. E., Ho L. C., 2004, ApJ, 610, 722

\bibitem[Greene \& Ho (2007)]{2007ApJ...670..92G} Greene J. E., Ho L. C., 2007, ApJ, 670, 92

\bibitem[Greene, Ho \& Barth (2008)]{2008ApJ...688..159} Greene J. E., Ho L. C., Barth A. J., 2008, ApJ, 688, 159

\bibitem[Hua \& Titarchuk(1995)]{1995ApJ...449..188H} Hua X.-M., Titarchuk L., 1995, ApJ, 449, 188 

\bibitem[Jiang et al.(2014)]{2014ApJ...796..106J} Jiang Y. F., Stone J. M., Davis S. W., 2014, ApJ, 796, 106 

\bibitem[Jin et al.(2013)]{2013MNRAS.436.3173J} Jin C., Done C., Middleton M., Ward M., 2013, MNRAS, 436, 3173

\bibitem[\protect\citeauthoryear{Jin et al.}{2009}]{Jin09} Jin C., Done C., Ward M., Gierli\'{n}ski M., Mullaney J., 2009, MNRAS, 398, L16

\bibitem[Jin, Ward \& Done (2012b)]{2012MNRAS.422.3268J} Jin C., Ward M., Done C., 2012b, MNRAS, 422, 3268 (J12b)

\bibitem[Jin, Ward \& Done (2012c)]{2012MNRAS.425..907J} Jin C., Ward M., Done C., 2012c, MNRAS, 425, 907 (J12c)

\bibitem[Jin et al. (2012a)]{2012MNRAS.420.1825J} Jin C., Ward M., Done C., Gelbord J., 2012a, MNRAS, 420, 1825 (J12a)

\bibitem[Kalberla et al.(2005)]{2005A&A...440..775K} Kalberla P. M. W., Burton W. B., Hartmann D., Arnal E. M., Bajaja E., Morras R., P\"{o}ppel W. G. L., 2005, A\&A, 440, 775 

\bibitem[Kelly et al.(2013)]{2013ApJ...779..187K} Kelly B. C., Treu T., Malkan M., Pancoast A., Woo J. H., 2013, ApJ, 779, 187 

\bibitem[Kaspi et al. (2000)]{2000ApJ...533..631} Kaspi S., Smith P. S., Netzer H., Maoz D., Jannuzi B. T., Giveon U., 2000, ApJ, 533, 631

\bibitem[Kunth, Sargent \& Bothun (1987)]{1987AJ.....93...29K} Kunth D., Sargent W. L. W., Bothun G. D., 1987, AJ, 93, 29

\bibitem[Laor(1991)]{1991ApJ...376...90L} Laor A., 1991, ApJ, 376, 90 

\bibitem[Laor et al.(1997)]{1997ApJ...477...93L} Laor A., Fiore F., Elvis M., Wilkes B. J., McDowell J. C., 1997, ApJ, 477, 93 

\bibitem[\protect\citeauthoryear{Leighly}{1999}]{Leighly99}Leighly K. M., 1999, ApJS, 125, 317

\bibitem[Ludlam et al.(2015)]{2015MNRAS.447.2112L} Ludlam R. M., Cackett E. M., G{\"u}ltekin K., Fabian A. C., Gallo L., Miniutti G., 2015, MNRAS, 447, 2112 

\bibitem[Magdziarz et al.(1998)]{1998MNRAS.301..179M} Magdziarz P., Blaes O. M., Zdziarski A. A., Johnson W.  N., Smith D. A., 1998, MNRAS, 301, 179

\bibitem[Makishima et al.(1986)]{1986ApJ...308..635M} Makishima K., Maejima Y., Mitsuda K., Bradt H. V., Remillard R. A., Tuohy I. R., Hoshi R., Nakagawa M., 1986, ApJ, 308, 635 

\bibitem[Marconi et al.(2008)]{2008ApJ...678..693M} Marconi A., Axon D. J., Maiolino R., Nagao T., Pastorini G.,  Pietrini P., Robinson A., Torricelli G., 2008, ApJ, 678, 693

\bibitem[Marconi et al. (2009)]{2009ApJ...698L.103M} Marconi A., Axon D. J., Maiolino R., Nagao T., Pietrini P., Risaliti G., Robinson A., Torricelli G., 2009, ApJ, 698, L103 

\bibitem[Markowitz et al.(2003)]{2003ApJ...598..935M} Markowitz A., Edelson R., Vaughan S., 2003, ApJ, 598, 935

\bibitem[Matt et al.(2014)]{2014MNRAS.439.3016M} Matt G., et al., 2014, MNRAS, 439, 3016 

\bibitem[McHardy et al.(2004)]{2004MNRAS.348..783M} M$^{c}$Hardy I. M., Papadakis I. E., Uttley P., Page M. J., Mason K. O., 2004, MNRAS, 348, 783

\bibitem[\protect\citeauthoryear{Middleton et al.}{2009}]{Middleton09}Middleton M., Done C., Ward M., Gierli\'{n}ski M., Schurch N., 2009, MNRAS, 394,250

\bibitem[Miniutti et al.(2009)]{2009MNRAS.394..443M} Miniutti G., Ponti G., Greene, J. E., Ho L. C., Fabian A. C., Iwasawa K., 2009, MNRAS, 394, 443

\bibitem[Mitsuda et al.(1984)]{1984PASJ...36..741M} Mitsuda K., et al., 1984, PASJ, 36, 741 

\bibitem[Morrison \& McCammon(1983)]{1983ApJ...270..119M} Morrison R., McCammon D., 1983, ApJ, 270, 119 

\bibitem[Nandra et al.(2007)]{2007MNRAS.382..194N} Nandra K., O'Neill P. M., George I. M., Reeves, J. N., 2007, MNRAS, 382, 194

\bibitem[Nowak et al.(1999)]{1999ApJ...515..726N} Nowak M. A., Wilms J., Vaughan B. A., Dove J. B., Begelman M. C., 1999, ApJ, 515, 726 

\bibitem[\protect\citeauthoryear{Osterbrock \& Pogge}{1985}]{Osterbrock85} Osterbrock D. E., Pogge R. W., 1985, ApJ, 297, 166

\bibitem[Rafter et al. (2011)]{2011ApJ...741..66R} Rafter S. E., Kaspi S., Behar E., Kollatschny W., Zetzl M., 2011, ApJ, 741, 66

\bibitem[Peterson(1997)]{1997iagn.book.....P} Peterson B. M., 1997, An introduction to active galactic nuclei, Publisher: Cambridge, New York Cambridge University Press, 1997 Physical description xvi, 238 p.~ISBN 0521473489 

\bibitem[\protect\citeauthoryear{Peterson et al.}{2004}]{Peterson04}Peterson B. M., et al., 2004, ApJ, 613, 682

\bibitem[\protect\citeauthoryear{Peterson et al.}{2005}]{Peterson05}Peterson B. M., et al., 2005, ApJ, 632, 799

\bibitem[Polletta et al.(2007)]{2007ApJ...663...81P} Polletta M., et al., 2007, ApJ, 663, 81 

\bibitem[Ponti et al.(2012)]{2012A&A...542A..83P} Ponti G., Papadakis I., Bianchi S., Guainazzi M., Matt G., Uttley P., Bonilla N. F., 2012, A\&A, 542, A83

\bibitem[S{\c a}dowski et al.(2015)]{2015MNRAS.447...49S} S{\c a}dowski  A., Narayan R., Tchekhovskoy A., Abarca D., Zhu Y., McKinney J. C., 2015, MNRAS, 447, 49 

\bibitem[Schurch \& Done(2007)]{2007MNRAS.381.1413S} Schurch N. J., Done C., 2007, MNRAS, 381, 1413 

\bibitem[Sutton et al.(2015)]{2015arXiv150301711S} Sutton A. D, Roberts T. P., Gladstone J. C., Walton D. J., 2015, arXiv:1503.01711 

\bibitem[Titarchuk(1994)]{1994ApJ...434..570T} Titarchuk L., 1994, ApJ,  434, 570 

\bibitem[Titarchuk \& Lyubarskij(1995)]{1995ApJ...450..876T} Titarchuk L., Lyubarskij Y., 1995, ApJ, 450, 876 

\bibitem[Uttley et al.(2014)]{2014A&ARv..22...72U} Uttley P., Cackett E. M., Fabian A. C., Kara E., Wilkins D. R., 2014, A\&ARv, 22, 72 

\bibitem[Vaughan et al.(2003)]{2003MNRAS.345.1271V} Vaughan S., Edelson R., Warwick R. S., Uttley P., 2003, MNRAS, 345, 1271

\bibitem[Wang et al.(2014)]{2014ApJ...793..108W} Wang J. M., et al., 2014, ApJ, 793, 108

\bibitem[\protect\citeauthoryear{Wilkinson \& Uttley}{2009}]{Wilkinson09}Wilkinson T., Uttley P., 2009, MNRAS, 397, 666

\bibitem[Zampieri \& Roberts(2009)]{2009MNRAS.400..677Z} Zampieri L., Roberts, T. P., 2009, MNRAS, 400, 677 

\bibitem[Zdziarski et al.(1996)]{1996MNRAS.283..193Z} Zdziarski A. A., Johnson W. N., Magdziarz P., 1996, MNRAS, 283, 193 

\bibitem[Zhang \& Wang(2006)]{2006ApJ...653..137Z} Zhang E. P., Wang J. M., 2006, ApJ, 653, 137 

\bibitem[Zhou et al.(2010)]{2010ApJ...710...16Z} Zhou X. L., Zhang S. N., Wang D. X., Zhu L., 2010, ApJ, 710, 16 

\bibitem[Zoghbi et al.(2010)]{2010MNRAS.401.2419Z} Zoghbi A., Fabian A. C., Uttley P., Miniutti G., Gallo L. C., Reynolds C. S., Miller J. M., Ponti G., 2010, MNRAS, 401, 2419 

\bibitem[{\.Z}ycki et al.(1999)]{1999MNRAS.309..561Z} {\.Z}ycki P. T., Done C., Smith D. A., 1999, MNRAS, 309, 561 

\end{thebibliography}
\end{document}